\documentclass[
  11pt,          % Schriftgröße
  3p,            % 3-Column Layout (ähnlich Journal Layout)
  review,        % Review-Modus (doppelte Zeilenabstände, für Review)
]{elsarticle}
\usepackage[utf8]{inputenc}
\usepackage{hyperref}
\usepackage{amsmath}
\usepackage{amssymb}
\usepackage{bm}
\usepackage{pgfplots}
\usepackage{tikz}
\usepackage{caption}
\usepackage{subcaption}
\usepackage{dutchcal}
\usepackage{upgreek}
\usepackage{todonotes}
\usepackage{xcolor}
\usepackage{siunitx}
\usepackage{graphicx}
\usepackage{float}
\usepackage{pgfplotstable}
\usepackage{lineno}
\usepgfplotslibrary{statistics} % <-- das ist wichtig!
\linenumbers
\usetikzlibrary{external}
\pgfplotsset{compat=1.18}

\usetikzlibrary{arrows,shapes}
\usetikzlibrary{arrows.meta}
\renewcommand{\paragraph}[1]{{\noindent\textit{#1:}}}

\newcommand{\diff} [1]{\mathrm{d}{#1}}
\newcommand{\grad}{\nabla}    %{\operatorname{grad}}%
\renewcommand{\div}{\nabla \cdot}%{\operatorname{div}}%
\newcommand{\intVol}[2]{\int_{#1} #2 \; \diff{V}}

\newcommand{\pdiffFrac}[2]{\frac{\partial #1}{\partial #2}}
\newcommand{\domain}{\mathit{\Omega}}
\newcommand{\fluidDomain}{\domain_\text{f}}
\newcommand{\solidDomain}{\domain_\text{s}}
\newcommand{\solventDomain}{\domain_\text{solvent}}
\newcommand{\airDomain}{\domain_\text{air}}

\newcommand\+{\mkern2mu} 

\newcommand{\forallHoldsText}[1]{\forall \+ #1 \colon \;}
\pgfplotscreateplotcyclelist{varyPaired}{
  {black},
  {gray},
  {dblue},
  {blue},
  {dred},
  {red},
  {dgreen},
  {green},
  {dorange},
  {orange},
}

\newcommand{\vek}{\boldsymbol}%{\vec}%

\newcommand{\field}{c}%{\psi}
\newcommand{\fielda}{c_{\text{a}}}%{\psi}
\newcommand{\fieldresult}{\Tilde{c}}
\newcommand{\sourcetermDeposit}{q_{\text{d}}}%{\psi}
\newcommand{\testFunc}{\psi}%{\mathit{\Psi}}%{\eta}%

\renewcommand{\div}{\nabla \cdot}%{\operatorname{div}}%
\newcommand{\diffFrac}[2]{ \frac{\diff{#1}}{\diff{#2}} }

\newcommand{\intSurf}[2]{\int_{#1} #2 \; \diff{S}}

\newcommand{\intX}[3]{\int_{#1}^{#2} #3 \; \diff{x}}
\newcommand{\xAverage}[1]{ \langle #1 \rangle_x }

\newcommand{\chr}[1]{ {#1}_\mathrm{c} }

\newcommand{\charLength}{\chr{l}}
\newcommand{\charVelocity}{\chr{u}}

\newcommand{\Dirac}{\updelta}

\newcommand{\indicator}[1][]{I_{#1}}

\newcommand{\SurfDirac}{\Dirac_S}

\newcommand{\normal}{\vek{n}}

\newcommand{\Ca}{\mathit{Ca}}  % capillary number
\newcommand{\Pe}{\mathit{Pe}}  % Peclét number
\usepgfplotslibrary{groupplots}
\newcommand{\volume}{V}

\newcommand{\pos}{\vek{x}}

\newcommand{\velocityScalar}{u}
\newcommand{\velocity}{\vek{\velocityScalar}}

\newcommand{\source}{q}
\newcommand{\boundaryflux}{g}
\newcommand{\diffusivity}{D}

\newcommand{\alphatilde}{\Tilde{\alpha}}
\newcommand{\betatilde}{\Tilde{\beta}}

\newcommand{\convGrad}[1]{ \pdiffFrac{#1}{t} + \velocity \cdot \grad #1 }

\newcommand{\steigungGradient}{m}
\newcommand{\facvis}{k^{\text{viscosity}}}
\newcommand{\facevap}{k^{\text{evap}}}
\newcommand{\facsigma}{k^{\text{surf}}}
\pgfplotscreateplotcyclelist{viridismylist}{
  {color={rgb,255:red,68; green,1; blue,84}},   % z.B. viridis[1]
  {color={rgb,255:red,59; green,82; blue,139}}, % z.B. viridis[2]
  {color={rgb,255:red,33; green,145; blue,140}},% z.B. viridis[3]
  {color={rgb,255:red,94; green,201; blue,97}}  % z.B. viridis[4]
}

\definecolor{dgreen}{HTML}{1b9e77}
\definecolor{dred}{HTML}{d95f02}
\definecolor{dblue}{HTML}{377eb8}
\definecolor{dorange}{HTML}{e6ab02}

\let\originaleps=\epsilon
\let\epsilon=\varepsilon
\let\varepsilon=\originaleps

\DeclareSIUnit\unitVelocity{\metre \per \second}
\DeclareSIUnit\unitVolume{\cubic\metre}
\DeclareSIUnit\unitDiff{\metre\squared \per \second}
\DeclareSIUnit\unitDensity{\kilogram \per \cubic\metre}
\DeclareSIUnit\unitDynVisc{\kilogram \per \metre \per \second}
\DeclareSIUnit\unitKinVisc{\metre\squared  \per \second}
\DeclareSIUnit\unitStress{\kilogram \per \metre \per \second\squared}
\DeclareSIUnit\unitSurfaceTension{\kilogram \per \second\squared}
\DeclareSIUnit\unitBodyForce{\metre \per \second\squared}
\DeclareSIUnit\unitEnergy{\kilogram\metre\squared \per \second\squared}
\DeclareSIUnit\unitEnergyDensity{\kilogram \per \metre \per \second\squared}
\DeclareSIUnit\unitMobilityCH{\cubic\metre \second \per \kilogram}
\DeclareSIUnit\unitMobilityACNestler{ \metre\squared \second \per \kilogram}
\DeclareSIUnit\unitMobilityAC{ \metre \second \per \kilogram}
\DeclareSIUnit\unitMobilityACtau{\kilogram \per \second \per \metre\squared}
\DeclareSIUnit\unitLaplaceCoeff{\kilogram \metre \per \second\squared}
\DeclareSIUnit\unitPotentialCoeff{\kilogram \per \metre \per \second\squared}

\pgfplotsset{
    colormap={viridis_rev}{
        rgb(0cm)=(0.9922,0.9059,0.1451);
        rgb(1cm)=(0.6667,0.8627,0.1961);
        rgb(2cm)=(0.3608,0.7843,0.3882);
        rgb(3cm)=(0.1529,0.6784,0.5059);
        rgb(4cm)=(0.1294,0.5647,0.5529);
        rgb(5cm)=(0.1725,0.4431,0.5569);
        rgb(6cm)=(0.2314,0.3176,0.5451);
        rgb(7cm)=(0.2784,0.1725,0.4784);
        rgb(8cm)=(0.2667,0.0039,0.3294);
    }
}
\pgfplotsset{
    colormap={inferno}{
        rgb(0cm)=(0.0015,0.0005,0.0139);
        rgb(1cm)=(0.230,0.018,0.109);
        rgb(2cm)=(0.423,0.079,0.179);
        rgb(3cm)=(0.635,0.208,0.274);
        rgb(4cm)=(0.823,0.321,0.375);
        rgb(5cm)=(0.956,0.518,0.451);
        rgb(6cm)=(0.992,0.688,0.380);
        rgb(7cm)=(0.998,0.857,0.334);
        rgb(8cm)=(0.988,0.998,0.644);
    }
}

\pgfplotsset{
    colormap={inferno_rev}{
        rgb(0cm)=(0.988,0.998,0.644);
        rgb(1cm)=(0.998,0.857,0.334);
        rgb(2cm)=(0.992,0.688,0.380);
        rgb(3cm)=(0.956,0.518,0.451);
        rgb(4cm)=(0.823,0.321,0.375);
        rgb(5cm)=(0.635,0.208,0.274);
        rgb(6cm)=(0.423,0.079,0.179);
        rgb(7cm)=(0.230,0.018,0.109);
        rgb(8cm)=(0.0015,0.0005,0.0139);
    }
}

\pgfplotstableset{
    create on use/y_scaled/.style={
        create col/expr={
            \thisrowno{1}/104.52*50000
        }
    }
}

\pgfplotstableset{
    create on use/y_scaled_gradient/.style={
        create col/expr={
            \thisrowno{2}/104.52*50000*\thisrowno{1}
        }
    }
}

\pgfplotstableset{
    create on use/y_gradient/.style={
        create col/expr={
            \thisrowno{3}
        }
    }
}

\pgfplotstableset{
    create on use/y_gradienttwo/.style={
        create col/expr={
            \thisrowno{2}
        }
    }
}

\pgfplotstableset{
    create on use/y_gradientone/.style={
        create col/expr={
            \thisrowno{1}
        }
    }
}

\pgfplotstableset{
    create on use/x_ca/.style={
        create col/expr={
            \thisrowno{0}
        }
    }
}

\pgfplotstableset{
    create on use/x_re/.style={
        create col/expr={
            \thisrowno{5}
        }
    }
}

\newcommand{\ComputeStatsFromFile}[4]{%
    \pgfplotstableread[col sep=space, header=false]{#1}\datatabletmp
    \pgfmathsetmacro{\N}{0}%
    \pgfmathsetmacro{\sum}{0}%
    \pgfmathsetmacro{\sumsq}{0}%
    \pgfplotstableforeachcolumnelement{#2}\of\datatabletmp\as\y{%
        \pgfmathsetmacro{\N}{\N + 1}%
        \pgfmathsetmacro{\sum}{\sum + \y}%
        \pgfmathsetmacro{\sumsq}{\sumsq + \y*\y}%
    }%
    \pgfmathsetmacro{\tmpmean}{\sum/\N}%
    \pgfmathsetmacro{\tmpstd}{sqrt((\sumsq/\N) - (\tmpmean*\tmpmean))}%
    \expandafter\def\expandafter#3\expandafter{\tmpmean}%
    \expandafter\def\expandafter#4\expandafter{\tmpstd}%
}

\usepackage{fancyhdr}

\makeatletter
\def\ps@pprintTitle{%
 \let\@oddhead\@empty
 \let\@evenhead\@empty
 \let\@oddfoot\@empty
 \let\@evenfoot\@empty
}
\makeatother

\usepackage{lineno}
\nolinenumbers
\begin{document}

\begin{frontmatter}
\title{Pore-scale modeling of capillary-driven binder migration during battery electrode drying}

\author[affil_1,affil_2]{Marcel Weichel}
\ead{marcel.weichel@kit.edu}

\author[affil_2,affil_3]{Martin Reder}

\author[affil_2]{Gerit Mühlberg}

\author[affil_4]{David Burger}

\author[affil_4]{Philip Scharfer}
\author[affil_4]{Wilhelm Schabel}

\author[affil_1,affil_2,affil_3]{Britta Nestler}

\author[affil_1,affil_2,affil_3]{Daniel Schneider}

\address[affil_1]{Institute of Nanotechnology (INT), Karlsruhe Institute of Technology, Hermann-von-Helmholtz-Platz 1, 76344 Eggenstein-Leopoldshafen, Germany}
\address[affil_2]{Institute for Applied Materials (IAM-MMS), Karlsruhe Institute of Technology, Strasse am Forum 7, 76131 Karlsruhe, Germany}
\address[affil_3]{Institute of Digital Materials Science (IDM), Karlsruhe University of Applied Sciences, Moltkestrasse 30, 76133 Karlsruhe, Germany}
\address[affil_4]{Thin Film Technology (TFT), Karlsruhe Institute of Technology (KIT), Strasse am Forum 7, 76131 Karlsruhe, Germany}

%\address[affil_4]{These authors contributed equally to this work.}

\begin{abstract}
Sodium-ion batteries employing hard carbon electrodes are considered a drop-in technology for lithium-ion batteries.
Electrode drying is a critical manufacturing step, as binder migration during pore emptying impacts the  mechanical integrity and electrical performance of the electrode. 
Existing modeling approaches predominantly rely  on the film shrinkage phase in a one dimensional way or neglect the capillary transport, resulting in a lack of physically consistent microstructure resolved predictions of binder migration.
In this work, a spatially resolved pore scale continuum model is extended to explicitly describe capillary driven binder transport during pore emptying. 
The model is applied to hard carbon microstructures with varying mean particle diameters. 
The simulations reveal that smaller particle sizes lead to a more homogeneous binder distribution, whereas higher evaporation rates and increased surface tension promote stronger binder gradients. 
Variations in solvent viscosity show only a minor influence on binder migration, as long as no hydrophilic or hydrophobic behavior is present.
Finally, the simulations demonstrate that an explicit description of capillary transport and microstructural effects is essential for accurately predicting binder migration and provides a basis for the targeted optimization of electrode drying processes.

% Natrium Ionen Batterienen mit ihren Hard Carbon elektrodengelten als drop in technolgie für Lithium ionen batterien. 
% Hierbei ist die Trocknung von Elektroden ein kritischer Schritt der durch migrationen von additiven während der Porenentleerung einen Einfluss auf die finalen mechanische und elektrische Eigenschaften hat. 
% Bis jetzt bestehende Modelle, die die Binder migration untersuchen, fokusieren sich auf 1D Modelle und die filmschrumpfungsphase oder lösen den kapillartransport während der porenentleerung nicht auf. 
% Hieraus ergibt sich eine fehelnde physikalisch Konsitende Beschreibung der Binder migration auf der Mikrostruktueben. 
% In dieser Arbeit wird ein spatially resolved pore scale contonnum model erweitert um eine differential gleichung welche die Binder migration auf grund von Kapillaren Kärften abbildet. 
% Dieses Modell wir auf Hard Carbon Mikrostrutkuren angewandt, welche sich in ihrem mittelen partikeldurchmesser unterscheiden. 
% Die Simulation zeigen, dass kleine Partikel eine homogenere Binderverteilung aufzeigen, sowie höhere Evapraotionsraten und Oberflächenspannungen einen stärkeren Gradienten der Bindervereilung haben. 
% Die Visksotät spielt bei den Simulationen einene untergeordente Rolle, welche durch die Beeinflussung des benetungswinkels aufgehoben wird. 
% Schließliocht zeigen die Simulationen, dass es für die binder migration notwenig ist den kapillartransport und die Mikrtroklsutr explizit zu modellieren. 
% Weiterhin ergeben sich durch die gezielte veränderung von Eigenschaften eine prozessoptimierte elektrodenherstellung.
\end{abstract}

\begin{keyword}
multiphase-field method, electrode drying, pore emptying, post-lithium battery systems, binder migration
\end{keyword}
\end{frontmatter}                 %Hauptteil
% \include{00-abstract}
   % Unterfiles sind schon angelegt
% \include{02-Stand-der-Forschung}          %können aber auch umbenannt werden.

% \include{01_Introduction} 
\section{Introduction}
The growing demand for high-energy, cost-efficient, and sustainable energy storage systems has intensified research on post lithium batteries, such as sodium-ion batteries.   
To fully exploit their electrochemical potential, the manufacturing process of battery electrodes need to be carefully reviewed. %of hard carbon~(HC)
This process involves several steps including mixing, coating, drying, calendering, and cutting.
Among these, the drying step is particularly critical, as even small variations can significantly alter the resulting microstructure and, consequently,  overall cell performance.
During drying, the solvent evaporates, and the solid framework, composed of active material, conductive carbon, and polymer binder forms the final microstructure, which influences electrode properties such as porosity, adhesion to the substrate, and particle cohesion.
Binder migration occurs, mainly during the pore-emptying stage of drying~\cite{burger2025simultaneous}, driven by capillary forces within the microstructure.
This migration can lead to a strongly inhomogeneous binder distribution, thereby affecting mechanical integrity, and  electronic conductivity.
Specifically, binder accumulation at the electrode surface reduces adhesion to the current collector~\cite{klemens2023process}.
% and increasing  interfacial resistance due to pore clogging.
Conversely, local binder depletion can compromise particle cohesion.
% Moreover, local binder accumulation can affect the electrical connectivity, ion transport pathways,
% and the wetting behavior during electrolyte infiltration.
Thus, drying remains a key step in achieving process reliability, as it defines process-structure-relations essential for mitigating binder migration. \\
% Thus, drying remains a key step in achieving both high energy density and process reliability, as it defines process-structure-relations essential for mitigating binder migration. \\
\paragraph{Physical understanding of solvent drying and binder migration}
In the recent decade, there has been a shift in the understanding of the physical processes underlying the drying process and the binder migration. 
Preliminary studies~\cite{jaiser2016investigation} hypothesized the formation of a dense consolidation layer, at which the active material and the binder accumulate.
Consequently, the idea of a top-down consolidation was proposed, suggesting that particle rearrangement and sedimentation dominate the drying behavior. 
Moreover, the aforementioned hypothesis posits that the binder migration was attributable to convective motion of the solvent, counteracted by molecular diffusion.
However, ex-situ imaging has demonstrated that particle distributions remain homogeneous during the drying process and that no consolidation front develops~\cite{jaiser2017microstructure}.
Therefore, the fundamental physical mechanism underlying binder migration is capillary-driven transport, which aligns well with experimental works on the optimization of the drying process by multi-stage drying profiles~\cite{jaiser2017development, altvater2024application}.
The binder moves in response to capillary pressure gradients between pores of different sizes.
Hence, it is evident that binder migration is directly coupled to the capillary solvent flow after reaching a consolidated enough structure. \\
\paragraph{Modeling approaches for electrode drying}
In addition to experimental approaches, modeling and simulation are key tools for understanding and optimizing different processes as well as determining process-structure-relations.
In the context of battery electrode drying, the distinct approaches can be categorized into homogenized continuum models~(HCM), coarse-grained molecular dynamics~(CGMD), discrete element methods~(DEM), and spatially resolved continuum models~(RCM).
HCM models~\cite{susarla2018modeling,stein2017mechanistic} reduce the drying process to a set of one-dimensional partial differential equations.
Thereby, these approaches establish a connection between macroscopic drying kinetics and process parameters.
However, a limitation of such approaches is that they neglect several physical processes, including pore-scale effects, such as capillary flow and wetting behavior.
Extensions of these approaches are made in~\cite{font2018binder,zihrul2023model,li2024numerical} by augmenting the governing equations with an additional convection-diffusion equation modeling binder transport.
These extended HCMs demonstrate a more pronounced binder gradient, under faster evaporation conditions.
Coarse-grained molecular dynamics~(CGMD) models describe the binder and carbon black as a single phase, which is termed as CBD phase. 
The process of solvent removal is modeled by the shrinkage of these phases, while the motion of the solvent is implicitly governed by the phenomenon of Brownian motion.
With these CGMD models, the microstructure and the binder distribution are captured~\cite{forouzan2016experiment,lombardo2022artistic,lombardo2022experimentally}, while pore-emptying physics cannot be modeled.
Discrete element method~(DEM) simulations focus on the initial drying stage and film shrinkage~\cite{lippke2023simulation}. 
There are also works that simulate the subsequent calendaring step using DEM~\cite{lippke2024drying}.
Furthermore, some studies investigate the final binder distribution employing simplified HCM models~\cite{zihrul2023model}.
The group of spatially resolved continuum models capture the movement of the fluid phase explicitly by tracking the liquid-gas interface during drying. 
Examples of this include the volume-of-fluid~(VOF) model by Wolf et al.~\cite{wolf2024computational} and the phase-field approach by Weichel et al.~\cite{weichel2025modeling}.
While the first study is restricted to spherical particles with no explicit representation of the surface wetting behavior, the latter is able to incorporate these physical mechanisms. 
A further class of approaches specifically modeling the pore-emptying process can be grouped into pore network models~(PNM) and continuum models~(CM). 
Continuum models describe the pore-emptying process at the macroscopic scale, where volume averaged quantities are employed to represent the underlying pore-scale behavior~\cite{whitaker1977simultaneous}. 
In contrast, pore network models rely on statistical representations of the pore network to investigate the influence of different parameters, such as viscosity~\cite{metzger2005influence,metzger2007isothermal,metzger2008viscous}.

\paragraph{Modeling approaches for binder migration during drying}
Most approaches to model binder migration are based on an early understanding of the drying process.
These approaches use a one-dimensional convection-diffusion equation to describe binder transport mostly during film shrinkage, while neglecting all capillary effects during pore emptying~\cite{font2018binder,zihrul2023model,li2024numerical}.
In these approaches, film shrinkage drives binder accumulation at the surface, a process that is partially compensated by diffusion, which transports binder towards the substrate.
These effects give rise to a dependence on the Peclét number, which correlates with the intensity of binder migration. 
However, it does not provide mechanistic insights as binder migration is driven by capillary transport that is not captured by the Peclet number.
Furthermore, the approach by Lombardo et al.~\cite{lombardo2021carbon} which is based on a coarse-grained molecular dynamics~(CGMD) model, enables the calculation of the final binder distribution. 
This model implicitly accounts for capillary forces by simulating interparticle forces of the CBD particles. 
Although existing modeling  frameworks yield valuable insights into drying kinetics and qualitative binder behavior, they generally fail to consistently capture capillary transport at the pore scale, which is the main mechanisms for  binder migration in porous electrodes.
Therefore, the present work aims to investigate drying dynamics and binder migration in wet film coated battery electrodes, using hard carbon (HC) as an example, by means of an extended modeling approach that explicitly resolves the underlying transport mechanisms. \\
\paragraph{Contribution of the present work} Building up on the existing model of~\cite{weichel2025modeling} for pore emptying, which directly resolves liquid–air interfaces, capillary forces, and wetting behavior within realistic electrode geometries, we introduce an additional system of partial differential equations to describe binder transport and deposition.
This extension enables a deeper physical understanding of the governing transport phenomena and supports the identification of improved electrode manufacturing strategies based on deeper insights into the influence of process parameters by the simulation.
The analysis links drying conditions, microstructural characteristics, pore-emptying dynamics, and the resulting binder distributions.
Therefore, we present the first spatially resolved continuum framework, that consistently captures capillary driven binder migration. 
Based on the previous work~\cite{weichel2025modeling}, the extended system of governing equations is presented in Sec.~\ref{sec:model}, which first summarizes the underlying framework and subsequently introduces the novel binder transport formulation.
The model is implemented in the multiphysics simulation framework PACE3D~\cite{hotzer2018parallel}, enabling parallelization via the Message Passing Interface (MPI) and allowing physics-based modeling of microstructural evolution driven by multiple physical mechanisms.
Sec.~\ref{sec:validation} validates the enhanced model using two representative test cases.
Section~\ref{sec:application} applies the model to experimental hard carbon microstructures to investigate the impact of various process parameters on the final binder distribution.
Finally, Sec.~\ref{sec:summary} concludes the paper and outlines directions for future research.

\section{Mathematical Formulation}
\label{sec:model}
To model the drying process in porous media such as battery electrodes, we consider a multiphase problem consisting of two fluids and multiple solid phases.
The fluid region is denoted by  $\fluidDomain$ and the solid region by $\solidDomain$, such that the full computational domain is $\domain = \fluidDomain \cup \solidDomain $.
Furthermore, the fluid domain consists of the solvent and air domain as $\fluidDomain = \airDomain \cup \solventDomain$.
The evaporation model of Weichel et al.~\cite{weichel2025modeling} is employed, which is briefly summarized in this section~\ref{sec:evaporation_model_sum}.
A detailed overview of the underlying drying framework may be found in previous publications~\cite{weichel2025modeling, reder2022phase}.
To describe binder migration, the governing system of partial differential equations is extended by an additional convection–diffusion equation as well as a deposition model in section~\ref{sec:binder_model_new}.
\subsection{Evaporation model}
\label{sec:evaporation_model_sum}
\paragraph{Phase evolution of a two-phase system}
The electrode geometry is represented using a phase-field method, where the interfaces between the phases are modeled by a smooth transition layer.
For a two-phase fluid system, the temporal evolution of the phase-field variable $\Tilde{\varphi}(\bm{x},t)$, representing the local volume fraction of one of the two phases at the spatial position $\bm{x}\in\fluidDomain$ and time $t$, can be described by an Allen–Cahn formulation of the form
\begin{align}
 \Dot{\Tilde{\varphi}}  =  M  (\betatilde \partial_{\Tilde{\varphi}}\psi - \alphatilde  \grad^2 \Tilde{\varphi}) , \qquad  \bm{x} \in \fluidDomain,
 \label{eq:allen-cahn}
\end{align}
where the boundary condition
\begin{align}
        \alphatilde \grad \Tilde{\varphi} \cdot \bm{n}^{\text{fs}} &= (\sigma_{\text{2s}}-\sigma_{\text{1s}})\partial_{\Tilde{\varphi}}h^{\text{ff}}(\Tilde{\varphi}), \qquad  \bm{x} \in \partial \fluidDomain
        \label{eq:wettingBC}
\end{align}
describes the surface wetting based on solid-fluid surface energies $\sigma_{\text{1s}}$ and $\sigma_{\text{2s}}$ according to Young's law, cf., e.g.~\cite{jacqmin1999calculation}.
Here, $\Dot{\Tilde{\varphi}}$ denotes the material derivative, $M$ the mobility, and $\grad^2 := \nabla \cdot \nabla$ the Laplacian operator.
The coefficients $\alphatilde$ and $\betatilde$ are defined as $\alphatilde = \sigma \varepsilon$ and $\betatilde = 18 \frac{\sigma}{\varepsilon}$, where $\sigma$ denotes the surface energy density between the two fluids and $\varepsilon$ controls the thickness of the diffuse interface.
The free-energy density $\psi(\Tilde{\varphi})$ is given by the double-well potential $\psi(\Tilde{\varphi}) = \Tilde{\varphi}^2(1-\Tilde{\varphi})^2$.
Since the phase fields satisfy the sum constraint, the second phase is implicitly defined by $1-\Tilde{\varphi}$, and only one evolution equation needs to be solved.
In addition, $\bm{n}^{\text{fs}}$ denotes the outward-pointing unit normal vector on the boundary of $\fluidDomain$, while the interpolation function $h^{\text{ff}}(\Tilde{\varphi}) = \Tilde{\varphi}^2(3-2\Tilde{\varphi})$ is used. 
% Contact angle conditions are governed by Young's law, where the solid-fluid surface energies are represented by $\sigma_{\text{2s}}$ and $\sigma_{\text{1s}}$.
Following the modifications proposed by Weichel et al.~\cite{weichel2025modeling} the evolution equation is extended to
\begin{align}
   \Dot{\Tilde{\varphi}} &= M\left[ \frac{36 \sigma}{\varepsilon }(2\Tilde{\varphi}^3 -3\Tilde{\varphi}^2 + \Tilde{\varphi})    - \sigma \varepsilon\left(  \grad^2 \Tilde{\varphi} - ||\grad\Tilde{\varphi}||\grad \cdot \bm{n} \right) \right] + v^{\text{e}} ||\grad \Tilde{\varphi}||, \qquad  \bm{x} \in \fluidDomain. 
    \label{eq:Allen-Cahn-Gleichung-Evap}
\end{align}
In this modification, the curvature minimizing dynamics of the Allen-Cahn equation are counteracted through the term $||\grad\Tilde{\varphi}||\grad \cdot \bm{n}$, thereby preventing an overestimation of surface tension effects, since these are already considered by a capillary term in the Navier-Stokes equations.
In addition, solvent evaporation is represented by the term $v^{\text{e}} ||\grad \Tilde{\varphi}||$, where the interface velocity $v^{\text{e}}$ is related to experimentally determined evaporation rates through the factor~$\kappa$ (unit \si{\per\second}) according to equation~\eqref{eq:kappa-and-evaprate}.\\
\paragraph{Fluid Motion}
To describe the motion of the two fluid phases, the incompressible Navier-Stokes equations of the form
\begin{subequations} \label{eq:Navier-Stokes}
  \begin{align}
       \rho \frac{\partial \boldsymbol{u}}{\partial t} + \rho \grad \boldsymbol{u}  \cdot \boldsymbol{u}&= -\grad p + \grad \cdot \left[ \mu \left( \grad\boldsymbol{u} + (\grad \boldsymbol{u})^{\text{T}} \right) \right] + \boldsymbol{K} +\rho \boldsymbol{f}_{\text{V}}, \label{eq:navier-stokes-moment}\\
    \grad \cdot \boldsymbol{u} &= 0 \label{eq:navier-stokes-cont}
  \end{align}
  \label{eq:nsg}
\end{subequations}
are employed within the fluid domain~$\fluidDomain$.
%
% Equations~\eqref{eq:navier-stokes-moment} and \eqref{eq:navier-stokes-cont} represent the momentum and continuity equations for incompressible flow of Newtonian fluids. 
Here, $\boldsymbol{u}$ denotes the velocity field, $\rho$ the mass density, $p$ the pressure, $\mu$ the dynamic viscosity, $\boldsymbol{K}$ the capillary term and $\boldsymbol{f}_{\text{V}}$ the external body forces acting on the system. 
In general, the quantities $\rho = \rho(\Tilde{\varphi})$, $\mu = \mu(\Tilde{\varphi})$ and $\boldsymbol{K}= \boldsymbol{K}(\Tilde{\varphi})$ depend on the phase-field variables. 
Furthermore, the capillary term is expressed as
\begin{align}
    \boldsymbol{K}(\Tilde{\varphi})
    =
    -\,\Tilde{\varphi}\,\nabla\Phi,
    \qquad
    \Phi
    =
    \betatilde\,\partial_{\Tilde{\varphi}}\psi
    -
    \alphatilde\nabla^{2}\Tilde{\varphi}.
\end{align}
% $   \boldsymbol{K}(\Tilde{\varphi}) = -\Tilde{\varphi} \grad \Phi$ with $\Phi =   \beta \partial_{\Tilde{\varphi}}\psi - \alpha  \grad^2 \Tilde{\varphi}$. \\
\paragraph{Extension to multiphase flow with rigid-body coupling}
To incorporate rigid-body interactions into the two-phase flow framework, a normalization procedure following Reder et al.~\cite{reder2022phase} is employed.
This allows for a diffuse representation of the fluid-solid interface and an extension of the equations~\eqref{eq:Allen-Cahn-Gleichung-Evap} and~\eqref{eq:Navier-Stokes} from the fluid domain~$\fluidDomain$ to the whole computational domain~$\domain$.
To this end, a multiphase formulation with $N$  phases $\alpha = 1,\ldots,N$ is introduced, employing phase-field variables $ \varphi_{\alpha}(\bm{x},t) \in [0,1]$ which represent the local volume fraction of phase $\alpha$. 
For a number of~$N^{\text{f}}$ fluid and~$N^{\text{s}}$ solid phases, we introduce the volume fractions of all fluid and solid phases as
\begin{align}
        \varphi^{\text{f}} = \sum_{\alpha=1}^{N^{\text{f}}}   \varphi_{\alpha}^{\text{f}} \qquad \text{and}  \qquad
            \varphi^{\text{s}} =  \sum_{\alpha=1}^{N^{\text{s}}}   \varphi_{\alpha}^{\text{s}} 
            \label{eq:normalized-phases}
\end{align}
yielding the summation constraint $\varphi^{\text{f}}+\varphi^{\text{s}} = 1$.
The normalization of the fluid phases yields 
\begin{align}
    \Tilde{\varphi}^{\text{f}}_{\alpha} := \frac{\varphi^{\text{f}}_{\alpha}}{\varphi^{\text{f}}}, \quad \text{for} \quad \varphi^{\text{f}}>0,
\end{align}
with the sum condition $\sum_{\alpha=1}^{N^{\text{f}}}  \Tilde{\varphi}^{\text{f}}_{\alpha} = 1 $.
For the case of two fluid phases, the system can be described by the single variable $\Tilde{\varphi} :=\Tilde{\varphi}^{\text{f}}_1 $, which is governed by equation~\eqref{eq:Allen-Cahn-Gleichung-Evap}.
Therefore, the multiphase problem is split into two-phase sub-problems with respect to $\{\varphi^{\text{s}},\varphi^{\text{f}}\}$ and $\{\Tilde{\varphi}_1,\Tilde{\varphi}_2\}$ within the fluid, cf.~\cite{reder2022phase}.
Furthermore, the boundary conditions at the diffuse fluid-solid interface are incorporated through the approach of Beckermann et al.~\cite{beckermann1999modeling} for the Navier-Stokes equations, and the approach of Li~\cite{li2009solving} for the Allen-Cahn equation. 
A detailed comparison of these approaches regarding the diffuse application of no-slip boundary conditions can be found in Reder et al.~\cite{reder2026benchmarks}. 
The application of these approaches gives rise to the coupled system of equations
\begin{subequations}
\label{eq:finales-gleichungssystem}
\begin{align}
  \begin{split}
    \partial_t(\rho h^{\text{fs}} \boldsymbol{u}) &= -\grad \cdot (\rho  h^{\text{fs}}  \boldsymbol{u}  \otimes\boldsymbol{u}) -h^{\text{fs}}\grad p \\
    &\qquad + \grad \cdot \left( h^{\text{fs}}\mu  \grad\boldsymbol{u}\right) + \boldsymbol{u} \grad \cdot (\mu \grad h^{\text{fs}}) +\rho h^{\text{fs}}  \boldsymbol{f}_{\text{V}} +  h^{\text{fs}}\boldsymbol{K}
  \end{split} \label{eq:1a} \\
  0 &= \grad \cdot (h^{\text{fs}}\boldsymbol{u}), \label{eq:1b} \\
  \begin{split}
    h^{\text{fs}}  \Dot{\Tilde{\varphi}} &=  M\Big[\betatilde h^{\text{fs}} \partial_{\Tilde{\varphi}} \psi - \alphatilde \left(\grad \cdot h^{\text{fs}} \grad \Tilde{\varphi} - h^{\text{fs}} ||\grad\Tilde{\varphi}||\grad \cdot \frac{\grad\Tilde{\varphi}}{||\grad\Tilde{\varphi}||}\right) \\
    &\qquad - \partial_{\varphi^{\text{f}}}h^{\text{fs}}||\grad\varphi^{\text{f}}|| (\sigma_{\text{2s}}-\sigma_{\text{1s}})\partial_{\Tilde{\varphi}}h^{\text{ff}}\Big] + h^{\text{fs}} v^{\text{e}} ||\grad \Tilde{\varphi}||
  \end{split} \label{eq:1c}
\end{align}
\end{subequations}
which are valid over the entire domain, i.e. $\bm{x} \in \domain$.
Herein, $\otimes$ denotes the dyadic product, defined as 
$(\boldsymbol{a}\otimes\boldsymbol{b})_{ij}=a_i b_j$.
Furthermore, the interpolation function between the fluid and solid interface is defined as  $h^{\text{fs}}(\varphi^\text{f}) = (\varphi^\text{f})^2(3-2\varphi^\text{f}) $ with $\varphi^\text{f}$ according to equation~\eqref{eq:normalized-phases}.
%%%

\subsection{Binder Transport Model}
\label{sec:binder_model_new}
In previous studies on binder migration during electrode drying~\cite{font2018binder, zihrul2023model, li2024numerical}, the redistribution of binder is commonly  described using  convection diffusion type transport equations. 
In particular, Zihrul et al.~\cite{zihrul2023model} employ a one-dimensional convection diffusion equation to model binder transport along the electrode thickness,
\begin{align}
\frac{\partial c}{\partial t} = D \frac{\partial^2c}{\partial^2 x} - c v,
\label{eq:binder-zihrul}
\end{align}
where, $c$ denotes the binder concentration, $D$ the diffusion coefficient, and $v$ the effective convective velocity. 
While such approaches successfully reproduce qualitative experimental trends~\cite{font2018binder, zihrul2023model, li2024numerical}, they are restricted to one-dimensional geometries and do not incorporate the capillary forces, which are the main reason for binder migration. 
To overcome these limitations, we introduce a two field formulation for binder migration, which distinguishes between mobile binder $\field$ dissolved in the solvent and deposited binder $\fielda$  during drying.
Binder transport is therefore described  by the following  three dimensional system of equations  
\begin{subequations} \label{eq:Binder-Equation-Start}
  \begin{alignat}{3}
    \dot{\field} := \convGrad{\field} &= \div \big(D \grad \field\big) - \sourcetermDeposit + \lambda, 
    \qquad && \bm{x} \in \solventDomain,
    \label{eq:Binder-eq-solvent-start}\\
    \frac{\partial \fielda}{\partial t} &= \sourcetermDeposit,
    \qquad && \bm{x} \in \airDomain
    \label{eq:Binder-eq-airt-start}.
  \end{alignat}
\end{subequations}
Here, $\sourcetermDeposit$ denotes a source term 
\begin{align}
   \sourcetermDeposit = \begin{cases}  \geq 0 \qquad \solventDomain \cup \airDomain \\ 0. \end{cases}
\end{align}
describing the irreversible deposition of binder from the solvent phase into the air phase.
Furthermore, a Lagrange multiplier $\lambda$ is introduced to enforce global binder mass conservation despite the continuous reduction of the solvent volume due to evaporation
\begin{align}
     \diffFrac{}{t}V_{\text{solvent}} =
     \diffFrac{}{t} \intVol{\solventDomain}{}
     \neq 0.
\end{align}
% \textcolor{red}{In physical terms, this term compensates for the apparent loss of binder concentration caused by solvent evaporation, thereby ensuring consistency in the overall mass balance.
% Furthermore, binder migration is resolved in three spatial dimensions, enabling a consistent approach to complex pore geometries}
% Unlike earlier models, which primarily associate binder redistribution with film shrinkage, the current formulation explicitly links binder transport to the evolving flow field and capillary-driven solvent motion. 
% \todo[inline]{Überarbeiten, mit hilfe von chatgpt war keine umformulierung, eventuell zu viel}
Regarding the boundary condition a homogeneous no-flux boundary condition is applied on the fluid boundary $\solventDomain$,
\begin{align}
   (D\grad \field) \cdot \normal &= 0, 
\qquad \bm{x} \in \partial\solventDomain. 
\label{eq:bc-binder}
\end{align}
Coupling the transport equations~\eqref{eq:Binder-Equation-Start} with the equation system~\eqref{eq:finales-gleichungssystem} ensures that capillary-driven binder migration is captured consistently, thereby overcoming earlier approaches that mainly associated binder redistribution with film shrinkage rather than capillary forces~\cite{font2018binder, zihrul2023model, li2024numerical}.
In addition, instead of a one-dimensional calculation of binder migration, a generally three-dimensional approach is employed.
To incorporate the no-flux boundary condition in a diffuse interface framework, 
an indicator function  $\indicator$ is introduced to distinguish the solvent domain from the surrounding phases, thereby following the approach of Li et al.~\cite{li2009solving}.
Using this indicator function, the binder transport equation~\eqref{eq:Binder-eq-solvent-start} can be reformulated to the whole-domain formulation
\begin{align}
\pdiffFrac{\field \indicator}{t} = - \indicator \bm{u} \cdot \grad \field 
+ \div \big( \indicator \diffusivity \grad \field \big) 
- \indicator \sourcetermDeposit + \lambda 
+ \field \pdiffFrac{\indicator}{t}, \qquad  \bm{x} \in \domain, 
\label{eq:final-cI}
\end{align}
where the last term accounts for the motion of the interface between solvent and air.
The derivation of equation~\eqref{eq:final-cI} can be found in detail in~\ref{sec:binder-derivation}.
A diffuse interface model for equation~\eqref{eq:final-cI} is obtained by approximating the indicator function as $\indicator \approx \Tilde{\varphi}$ using the normalised phase variable~$\Tilde{\varphi}$ of the solvent.
To isolate the individual effects of evaporation and binder deposition, two limiting cases are considered before discussing the fully coupled system.
Based on these limiting cases, the corresponding validation setups are defined in section~\ref{sec:validation}. \\
\paragraph{Special case without deposition and evaporation}  
In the absence of evaporation, the solvent volume remains constant and
\begin{align}
    \diffFrac{}{t} V_{\text{solvent}} =
    \diffFrac{}{t} \intVol{\solventDomain}{} = 0
\end{align}
holds.
Consequently, the Lagrange multiplier vanishes, $\lambda = 0$.
Furthermore, if binder deposition into the air phase is neglected, the deposition source term can be set to $\sourcetermDeposit = 0$.
Under these assumptions, the binder transport equation~\eqref{eq:final-cI} reduces to a purely convection-diffusion equation in the form of 
\begin{align}
\pdiffFrac{\field \indicator}{t} = - \indicator \bm{u} \cdot \grad \field 
+ \div \big( \indicator \diffusivity \grad \field \big)  
+ \field \pdiffFrac{\indicator}{t}. 
\label{eq:final-cI-no-deposition-no-lagrange}
\end{align}
This limiting case is used to validate the implementation by means of the bubble rise benchmark proposed by Reder et al.~\cite{reder2024viscous} in section~\ref{sec:validation-bubble}, where neither evaporation nor binder deposition is present and therefore, the binder concentration is transported according to the solvent motion and completely remains in the moving solvent domain. \\
\paragraph{Special case without deposition}
In a second step, binder evolution is investigated in the presence of evaporation, while deposition into the air phase is still neglected.
Accordingly, the deposition source term is set to $ \sourcetermDeposit = 0$, whereas the Lagrange multiplier can no longer be omitted, since
\begin{align}
    \diffFrac{}{t} V_{\text{solvent}} =
    \diffFrac{}{t} \intVol{\solventDomain}{} \neq 0 
\end{align} 
holds due to solvent evaporation.
To ensure global binder mass conservation, the constraint
\begin{equation}
    \diffFrac{}{t} %\left(
    \intVol{\domain}{\field\indicator }
    %- \intVol{\domain}{ (1-\indicator) \fielda} \right)
    = 0,
    \label{eq:constraint_without_deposition}
\end{equation}
must be fulfilled.
Therefore,  the Lagrange multiplier $\lambda$ is defined as 
\begin{align}
    \lambda= w_{\text{c}} p_{\text{c}}, \quad
    w_{\text{c}} =\Tilde{\varphi} (1-\Tilde{\varphi}),
\end{align}
where $p_{\mathrm{c}}$ is a global amplitude obtained through a projection step and $w_{\text{c}}$ a distribution function to re-distribute missing binder associated with evaporated solvent over the fluid-fluid interface.
In this case, the binder concentration stays inside the solvent while being globally preserved according to equation~\eqref{eq:constraint_without_deposition}.
This configuration is investigated using the evaporation of a fluid phase in a capillary geometry, as described in section~\ref{sec:binder-migration-validation-evap}.
\\
\paragraph{General case with deposition}
In the general case, the concentration~$\fielda$ of deposited binder in the air is non-zero and the conservation constraint reads
\begin{equation}
    \diffFrac{}{t} \left(\intVol{\domain}{\field\indicator }
    - \intVol{\domain}{ (1-\indicator) \fielda} \right)
    = 0.
    \label{eq:constraint_with_deposition}
\end{equation}
The binder deposition is explicitly modeled by the source term~$\sourcetermDeposit$, which is assumed to depend on the local volume fraction and deposited binder concentration $\sourcetermDeposit =  \sourcetermDeposit(\varphi_1, \varphi_2,...,\varphi_N, \fielda(\bm{x},t))$.
It is modeled as 
\begin{align}
    \sourcetermDeposit = w_{\text{d}} p_{\text{d}}, \quad
    w_{\text{d}} =  \Tilde{\varphi} (1-\Tilde{\varphi}) \left( f \, \varphi^{\text{s}} + (1 - f)  \right),
    \label{eq:source-deposition}
\end{align}
where $f$ is a model parameter, that controls the deposition of binder at the triple phase regions, i.e. for $f = 0$, all binder will be deposited at the fluid-fluid interface while for $f=1$ the deposition is restricted to triple phase regions.
Therefore, the separation parameter~$f$ can be used to model increased deposition of binder near the solid particle. 
In the case of deposition in the fluid-fluid interface, the binder remains in the carbon binder domain, which fills the space between the particles and holds this part of the binder in the electrode.
The parameter $p_{\text{d}}$ defines the corresponding deposition amplitude. \\
\paragraph{Solving method} A split-step approach is used to solve the binder transport equations, which separates transport, deposition, and mass-conservation correction into consecutive substeps.
All model equations are discretized using a finite difference method on a Cartesian grid.
% The model is implemented in the multiphysics simulation framework PACE3D~\cite{hotzer2018parallel} enabling parallelization via Message Passing Interface.
The corresponding algorithm to update the binder field, which is executed for each time step~$n$, is subsequently described.
Therefore, the information about $\indicator^{n+1}$ is known a priori to the concentration calculation.
\begin{enumerate}
   \item Solving the transport equation~\eqref{eq:final-cI} without the source term to get the interim solution:
    \begin{equation}
        (\field \indicator)^* = (\field \indicator)^n 
        + \Delta t \Big[ - \indicator \bm{u} \cdot \grad \field 
        + \div (\indicator \diffusivity \grad \field) \Big]^n
        + \field \left(\indicator^{n+1} - \indicator^n\right).
    \end{equation}
    \item Calculate the difference of initial binder mass in the solvent and the binder mass in the solvent of the interim solution
    \begin{align}
    \Delta_{cI} &=  \displaystyle \intVol{\domain}{ \field \indicator (\bm{x},t^0) }
        - \displaystyle \intVol{\domain}{ (\field \indicator)^*(\bm{x},t^n) }.
    \end{align}
    \item Update the binder concentration $\fielda$ in the air phase to model the binder deposition according to:
    \begin{align}
        \sourcetermDeposit &= w_{\text{d}} p_{\text{d}}, \\
        w_{\text{d}} &= f\, \Tilde{\varphi} (1-\Tilde{\varphi}) \varphi^{\text{s}} 
        + (1-f)\, \Tilde{\varphi} (1-\Tilde{\varphi}), \\
        p_{\text{d}} &= 
        \frac{
         \Delta_{cI}
        - \displaystyle \intVol{\domain}{ (1-\indicator^n) \fielda(\bm{x},t^n) }
        }{
        \displaystyle \intVol{\domain}{ w_{\text{d}}(\bm{x},t) } \Delta t
        }, \\
        (\fielda)^{n+1} &= (\fielda)^n + \sourcetermDeposit.
    \end{align}
    % with $\omega$ as a parameter between $\omega \in [0,1]$, whereas if $\omega = 1$ all the binder, which disappears due to evaporation, will be redistributed. 
    \item Ensure mass conservation by the correction step to obtain the new solution for $\field \indicator$:
    \begin{align}
        \lambda &= w_{\text{c}}\, p_{\text{c}}, \\
        w_{\text{c}} &= \Tilde{\varphi} (1-\Tilde{\varphi}), \\
        p_{\text{c}} &= 
        \frac{
         \Delta_{cI}
        - \displaystyle \intVol{\domain}{ (1-\indicator^{n+1}) c_{\text{a}}(\bm{x},t^{n+1}) }
        }{
        \displaystyle \intVol{\domain}{ w_{\text{c}}(\bm{x},t) } \Delta t 
        }  \\
        (\field \indicator)^{n+1} &= (\field \indicator)^* + \lambda.
    \end{align}
\end{enumerate}
\section{Validation}
\label{sec:validation}
In this section, the model is validated using two examples without deposition showing conservation of mass in the fluid despite movement of phase boundaries and evaporation. In the initial example, a liquid droplet is transported in a gravitational field to ensure that the initial amount of binder remains within the upwards moving liquid.
In the second example, a fluid undergoes evaporation, while the initial amount of binder should remain within the fluid. 
Additionally, the gradient can be balanced by diffusion of the concentration.
The results are evaluated based on the interpolated concentration
\begin{align}
    \fieldresult = \field I + (1-I)\fielda.
    \label{eq:fieldresult}
    \end{align}   
\subsection{Bubble rise problem}
\label{sec:validation-bubble}
\paragraph{Setup and parameters}  In order to investigate the conservation of mass in a moving fluid without evaporation, case~2 of the bubble rise problem from~\cite{Hysing2009,Aland2012Benchmark} is utilized.
The simulation parameters are summarized in Table~\ref{tab:iniital-paramters-binder-bubble-rise}. 
For the purpose of this study, a spherical bubble of fluid 2 ($\varphi_2$) with a radius of $R=0.25$ and an initial concentration of $c_0 = 0.01$ is placed within the simulation domain.  
\begin{table}[htbp]
\centering
    \caption{Overview of the simulation parameter for bubble rise problem.}
% \begin{ruledtabular}
  \begin{tabular}{l l l l l l l l l}
   \hline
    \hline
   \text{} & $\rho_1$ & $\rho_2$ & $\mu_1$ & $\mu_2$ & $g$ & $\sigma$ & $\kappa$ & $D$ \\
    \text{} &in $\si{\kg\per\cubic\metre}$ & in $\si{\kg\per\cubic\metre}$ & in $\si{\pascal\second}$ & in $\si{\pascal\second}$ & in $\si{\metre\per\square\second}$ & in $\si{\newton\per\metre}$ & in $\si{\per\second}$ & in \si{\unitDiff} \\
    \hline  
   \text{} & $1000$ & $1$ & $10$ & $0.1$ & $0.98$ & $1.96$ & $0$ & $0$ \\
            \hline
    \hline
    \end{tabular}

    % \end{ruledtabular}
    \label{tab:iniital-paramters-binder-bubble-rise}
\end{table}

\begin{figure}[htbp]
  \centering
\includegraphics[]{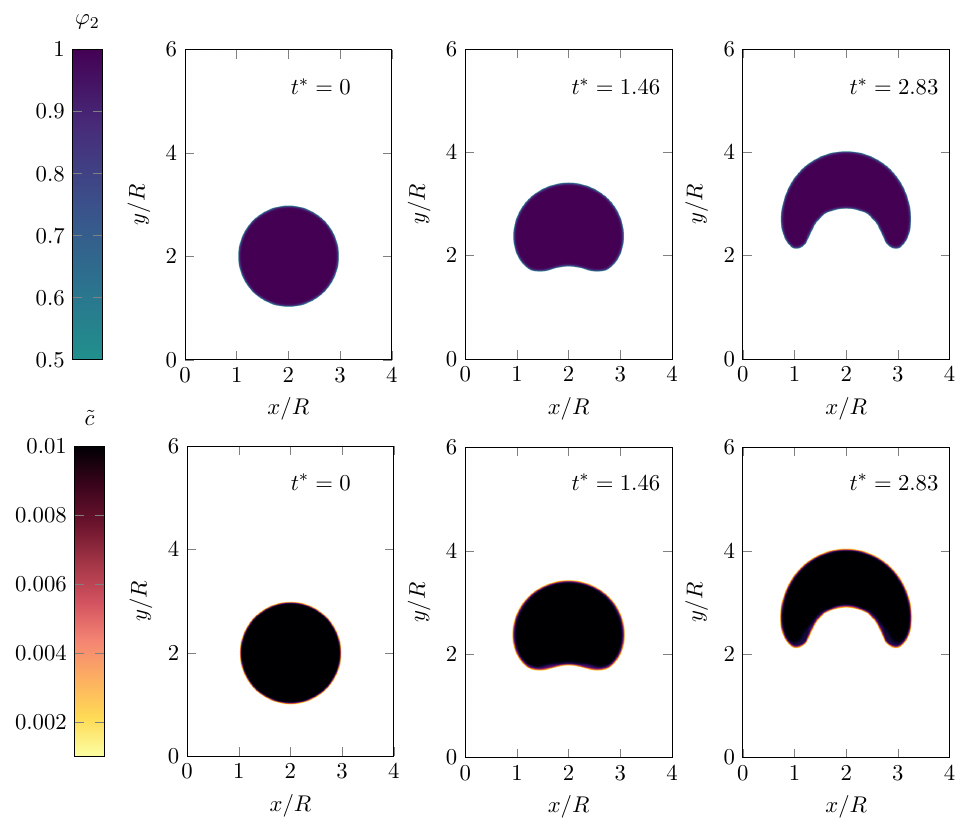}
  \caption{Spatial representation of the phase-field variable $\varphi_2$ (top) and the concentration $cI$ (middle) for the three dimensionless time points ($t^* = \sqrt{g/R}$): $t^* = 0$,  $t^* = 1.46$, and $t^* = 2.83$. %as well as the volume integral of the concentration $\fieldresult$ over the dimensionless time  (bottom).
  }
  \label{fig:bubble-rise-result}
\end{figure}

\paragraph{Results} Considering the dimensionless time $t^* = \sqrt{g/R}$, figure~\ref{fig:bubble-rise-result} shows the results of the phase field variable $\varphi_2$ (top)  and the concentration distribution $\fieldresult$ (middle) as a function of the geometric dimensions for the three points in time  $t^* = 0$, $t^* = 1.46$, and $t^* = 2.83$.
% Furthermore, the  
%  is depicted as a function of the dimensionless time $t^* = \sqrt{g/R}$ (bottom).
$\varphi_2$ undergoes a transformation over time, evolving from a spherical to a non-convex shape. 
This behavior is in accordance with the findings reported in Reder et al.~\cite{reder2024viscous}.
In addition, the concentration distribution $\fieldresult$ exhibits a similar tendency, manifesting analogous shapes over time. 
Furthermore, the total amount of binder, quantified by the volume integral  of the concentration distribution $\fieldresult$,
\begin{equation*}
    \intVol{\domain}{\fieldresult(\bm{x},t) } = 1.86\cdot 10^{-3}
\end{equation*}
is conserved over time.
Thus, the model is capable of carrying the binder within a moving fluid and remaining a constant mass and concentration, as can be seen in Figure~\ref{fig:bubble-rise-result} (bottom), where the volume integral displays a constant value over the specified period.  

\subsection{Binder migration during evaporation in a two fluid system}
\label{sec:binder-migration-validation-evap}
\begin{figure}[htbp]
   \centering
       \includegraphics[scale =0.5]{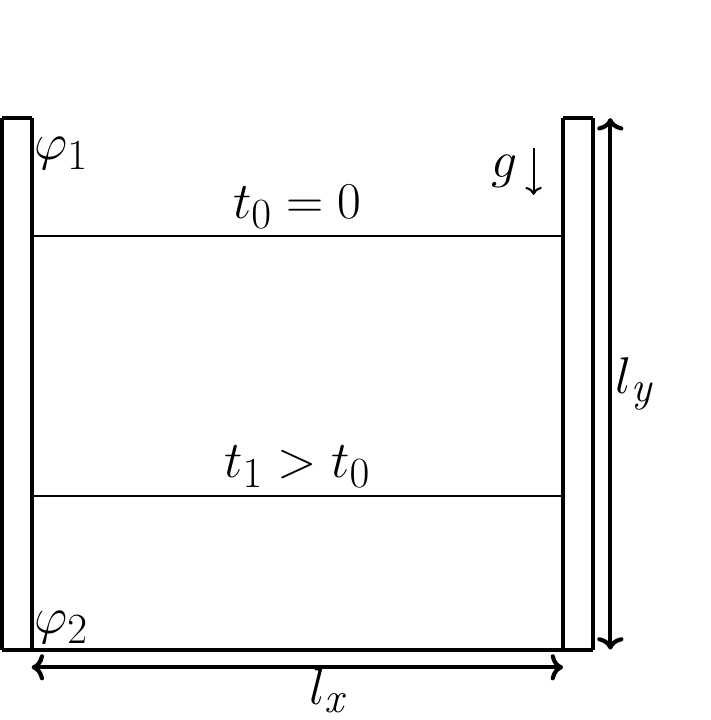}
       \caption[]{Schematic representation of the validation setup for the binder evolution in a drying fluid.}
       \label{fig:Initial setup-Riser}
   \end{figure}

\begin{figure}[htbp]
    \centering
    % Erste Subfigure
    \begin{subfigure}[b]{0.49\textwidth}
    \centering
 \includegraphics[]{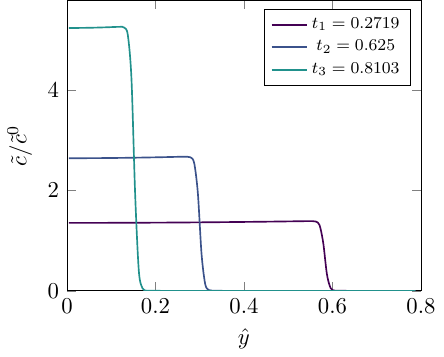}
    \end{subfigure}
    \hfill
    % Zweite Subfigure
      \begin{subfigure}[b]{0.49\textwidth}
    \centering
 \includegraphics[]{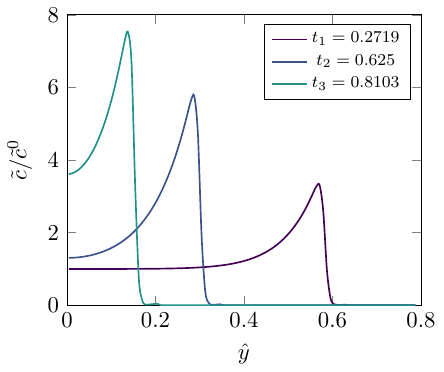}
    \end{subfigure}
    \caption{Normalized concentration profiles $\fieldresult/\fieldresult^0$ as a function of the height $\hat{y}$ at  $x = l_{\text{x}}/2$ at three dimensionless times: $t_1=\SI{0.2719}{}$, $t_2=\SI{0.625}{}$, $t_3=\SI{0.8103}{}$, and the P\'eclet numbers  $Pe = 12.5\cdot 10^{-2}$  (left) and $Pe = 12.5$ (right).}
    \label{fig:validation-two-fluid-evap}
\end{figure}

\paragraph{Setup and parameters} In order to investigate the behavior of binder migration in presence of evaporation, the validation case \textit{contact angle at a moving contact line} from~\cite{weichel2025modeling} is employed, with slightly modified parameters.
In contrast to the original setup, the contact angle is set to $\theta =\ang{90}$ instead of $\theta =\ang{60}$, which results in a horizontal drying front.  
A schematic sketch for the present simulation setup is shown in Fig.~\ref{fig:Initial setup-Riser}, while the corresponding simulation parameters are summarized in Table~\ref{tab:Initial parameters-binder-validation-2}. 
An initial dimensionless binder concentration of $c_0 = 0.01$ is chosen in the region of $\varphi_2$. 
The rates of evaporation are determined  by the  parameter $\kappa $.
% Utilizing the two  values $\kappa_1 = \SI{100}{\per\second}$ and $\kappa_2 = \SI{10000}{\per\second}$, as well as a diffusivity  of $D=\SI{1.579e-4}{\unitDiff}$ 
Furthermore, the system is characterized by the P\'eclet numbers $\Pe= 12.5\cdot 10^{-2}$ and $\Pe= 12.5$, respectively.
The P\'eclet number is thereby defined through 
\begin{equation}
    \Pe = \charVelocity\charLength/D = \kappa l_{\text{y}}^2/D.
\end{equation} 
This setup closely resembles the simulation setup used by Li et al.~\cite{li2024numerical}, who have employed the same P\'eclet numbers, therefore the diffusivity are fitted according to the characteristic velocity's.  \\
\begin{table}[htbp]
\centering
    \captionof{table}{Overview of the simulation parameter for the validation case contact angle at a moving contact line.}
% \begin{ruledtabular}
  \begin{tabular}{l l l l l l l l l l}
    \hline
    \hline
    & $l_{\text{x}}$ & $l_{\text{y}}$ & $\rho_1$ & $\rho_2$ & $\mu_1$ & $\mu_2$ & $g$ & $\sigma$  \\
    & in $\si{\micro\meter}$ & in $\si{\micro\meter}$ & in $\si{\kg\per\cubic\metre}$ & in $\si{\kg\per\cubic\metre}$ & in $\si{\pascal\second}$ & in $\si{\pascal\second}$ & in $\si{\metre\per\square\second}$ & in $\si{\newton\per\metre}$  \\
    \hline  
    & $56$ & $56$ & $1.225$ & $997$ & $1.72\cdot 10^{-5}$  & $1\cdot 10^{-3}$& $9.81$ & $1.72\cdot 10^{-5}$ \\
            \hline
    \hline
    \end{tabular}

    % \end{ruledtabular}
    \label{tab:Initial parameters-binder-validation-2}
\end{table}
\paragraph{Results} Figure~\ref{fig:validation-two-fluid-evap} displays the normalized concentration profiles $\fieldresult/\fieldresult^0$ as a function of the normalized height $\hat{y}$ at three dimensionless times $t_1=\SI{0.2719}{}$, $t_2=\SI{0.625}{}$, and $t_3=\SI{0.8103}{}$. 
The outcomes for $Pe = 12.5\cdot 10^{-2}$ are shown on the left of Figure~\ref{fig:validation-two-fluid-evap}, while those for $Pe = 12.5$ are shown on the right. 
Utilizing a diffuse interface leads to the formation of a transition zone subsequent to the physical liquid front, thereby elucidating the decline in the concentration curves depicted in Figure~\ref{fig:validation-two-fluid-evap}.
It can be seen that as the P\'eclet number increases, the rate of evaporation of the liquid accelerates. 
Furthermore, the differences between the maxima and minima are more pronounced for a high P\'eclet number than for a low P\'eclet number. 
This outcome is anticipated, given that convective transport dominates at a P\'eclet  number of $\Pe = 12.5$, while diffusive transport prevails at a P\'eclet number of $\Pe = 12.5\cdot 10^{-2}$. 
Furthermore,  a qualitative comparison  with the studies by Font et al.~\cite{font2018binder} and Li et al.~\cite{li2024numerical} reveals behavior consistent with the present results. 
Besides correctly capturing the dominant transport phenomena, these findings demonstrate that, if no deposition of binder into the air phase is considered, the proposed model reproduces the given results from literature.
\section{Application}
\label{sec:application}
In this section, the proposed model is applied to investigate capillary induced  binder migration in battery electrodes. 
The influence of key process and material parameters, like the solvent viscosity, evaporation rate and surface tension is systematically analyzed by varying these parameters. 
In addition, the impact of wettability is assessed by considering different contact angle conditions.
The resulting binder migration is quantified using a normalized vertical binder gradient, enabling a direct comparison between microstructures and process parameters.
% Thereby, the simulation uses real battery electrodes obtained from Klemens et al.~\cite{klemens2023process}, which have previously been analyzed for their drying behavior under slightly different segmentation and parameters settings in Weichel et al.~\cite{weichel2025modeling}. 

\subsection{Preprocessing and simulation setup}
\begin{figure}[htbp]
\centering
\begin{subfigure}[c]{0.5\textwidth}
\centering
\includegraphics[]{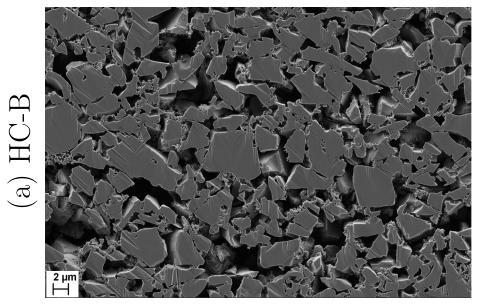}
\subcaption*{(i)}
\end{subfigure} 
\begin{subfigure}[c]{0.48\textwidth}
\centering
\includegraphics[]{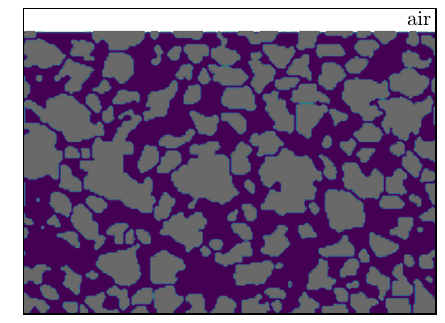}
\subcaption*{(ii)}
\end{subfigure} 
\begin{subfigure}[c]{0.50\textwidth}
\centering
\vspace{0.25cm}
\includegraphics[]{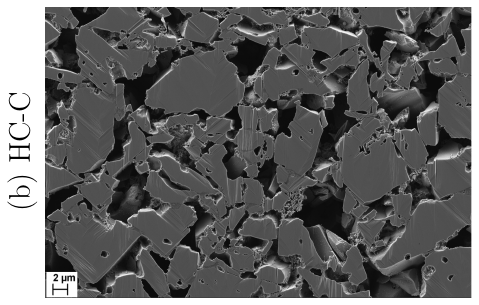}
\subcaption*{(i)}
 \label{fig:HC-C-mikro}
\end{subfigure} 
\begin{subfigure}[c]{0.48\textwidth}
\centering
\includegraphics[]{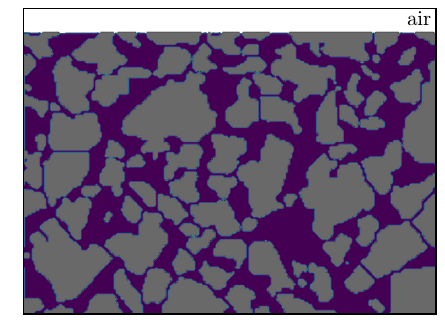}
\subcaption*{(ii)}
 \label{fig:HC-C-sim}
\end{subfigure} 
\caption{SEM images of dried microstructures (i) with different pore size distributions and particle diameters, labeled (a) HC-B and (b) HC-C~\cite{klemens2023process}. The corresponding digital representation of the initial three-phase microstructure (ii) composed of air (white), particles (gray), and binder (black) dissolved in the solvent.  The concentration of the binder is set to $c=0.01$.}
\label{fig:microstructures-initial}
\end{figure}

\paragraph{Microstructures}
The microstructures used in this study are obtained from the real hard carbon electrodes published by Klemens et al.~\cite{klemens2023process} and were  subsequently processed following the workflow described by Weichel et al.~\cite{weichel2025modeling}.
In the present work, the two considered microstructures are denoted as HC-B and HC-C, consistent with the terminology used in previous studies~\cite{klemens2023process,weichel2025modeling}.
Compared to Weichel et al.~\cite{weichel2025modeling}, the image segmentation parameters were slightly adjusted, leading to minor changes in the resulting pore size distributions.
These modifications do not alter the qualitative microstructural characteristics but influence quantitative measures such as breakthrough times.
HC-C generally exhibits a broader pore size distribution and a larger mean particle diameter compared to HC-B~\cite{klemens2023process}.
Both microstructures are characterized by the solid volume fractions~$\chi$ and the area-specific weight of the dry coating~$M$.
The solid volume fraction $\chi$ is obtained directly from the segmented simulation input, while the area-specific mass~$M$ is taken from experimental coating data.
For HC-B, these values are $\chi^{\text{HC-B}}_{\text{solid}}= 0.465$ and $M^{\text{HC-B}}_{\text{S}} = \SI{79.5e-3}{\kg\per\meter\squared}$, whereas HC-C is defined by $\chi^{\text{HC-C}}_{\text{solid}}= 0.549$ and $M^{\text{HC-C}}_{\text{S}} = \SI{78.4e-3}{\kg\per\meter\squared}$. 
It is to be noted, that the value for $\kappa$ is sensitive to the choice of $\chi$, which is influenced by the segmentation process.
The simulation domain represents a physical section of the microstructure with a width of $l_{\text{x}}=\SI{55.0}{\micro\meter}$ and a dry film thickness of $l_{\text{y}}=\SI{37.5}{\micro\meter}$.
A total physical simulation height of $l_{\text{y}} =\SI{40.07}{\micro\meter}$ is employed to account for the air phase. 
The initial binder concentration $c$ is set to $c=0.01$ as a dimensionless concentration, and uniformly initialized within the solvent filled regions.
Across the diffuse interface, the concentration is smoothly interpolated following the variable $\Tilde{\varphi}$.
Figure~\ref{fig:microstructures-initial} illustrates the original SEM pictures of the electrodes, the segmented simulation input, and the corresponding initial concentration.

\paragraph{Parameters}
Table~\ref{tab:InitialParametersHC} provides an overview of all physical parameters utilized in the simulations, including fluid densities, viscosities, surface tension, gravity, mobility, and the binder diffusion coefficient in the solvent.
To obtain the velocity of the receding liquid surface $v^{\text{e}}$ (see eq.~\ref{eq:Allen-Cahn-Gleichung-Evap}), we employ the parameter 
\begin{align}
 \kappa = \frac{\rho_{\text{solid}} \chi_{\text{solid}}}{\rho_{\text{solvent}} M_{\text{S}} } \Dot{r}_{\text{solvent}}, 
 \label{eq:kappa-and-evaprate}
\end{align}
which links the  prescribed evaporation rate of $\Dot{r}_0 =\SI{9}{\gram\per\square\metre\per\second}$ to the simulation framework~\cite{weichel2025modeling}.
Using this relation and the information about the microstructures, we obtain $\kappa^{\text{HC-B}} = \SI{0.1089}{\per\second} $ and $\kappa^{\text{HC-C}} =  \SI{0.1303}{\per\second}$ for HC-B and HC-C, respectively. 
Binder separation is modeled by the separation factor $f=0.9$.
Accordingly, \qty{90}{\percent} of the binder is deposited at the evolving triple phase boundaries, while the remaining \qty{10}{\percent} is assigned to the fluid-fluid interface.
This choice reflects experimentally observed preferential binder accumulation at particle surfaces, which is more pronounced in slurries with smaller particle diameters~\cite{klemens2023process}.
Furthermore, the computational grid consists of $n_x=\num{608}$ cells in the $x$-direction and $n_y=\num{448}$ cells in the $y$-direction.
Using the characteristic length $\charLength = l_{\text{x}} = \SI{55.0}{\micro\meter} $ the dimensionless spatial step sizes are $\Delta x^*=\Delta x/\charLength=1.64\cdot 10^{-3} $ and  $\Delta y^*=\Delta y/\charLength = 1.65\cdot 10^{-3}$.
\begin{table}[htbp]
\centering
    \captionof{table}{Overview of the simulation parameter for hard carbon microstructures for the present study.}
  \begin{tabular}{l c c c c c }
    \hline
    \hline
    parameter & symbol &  physic. value & unit \\
    \hline  
    % \hline
    physical length & $l_{\text{x}}$ & $55.0$ & \si{\micro\meter} \\
    physical height & $l_{\text{y}}$ & $40.07$& \si{\micro\meter} \\
    density hard carbon & $\rho_{\text{HC}}$ & $2062.0$ & \si{\kg\per\cubic\metre}  \\
    density  solvent & $\rho_{\text{solvent}}$ & $997.0$ & \si{\kg\per\cubic\metre}  \\
    density air & $\rho_{\text{air}}$ & $1.225$ & \si{\kg\per\cubic\metre} \\
    viscosity solvent & $\mu_{\text{solvent},0}$ & $1.0$ & \si{\pascal\second}\\
    viscosity air &$\mu_{\text{air}}$ & $1.72\cdot 10^{-5}$ & \si{\pascal\second}\\
    surface tension & $\sigma_0$ & $72.4\cdot 10^{-3}$ & \si{\newton\per\metre}  \\
    gravity & $g$ & $9.81$ & \si{\metre\per\square\second} \\
   phase-field mobility & $M$ & $30$ & \si{\unitMobilityAC} \\
    binder diffusion coefficient & $D$ &$1.12\cdot 10^{-16}$ & $\si{\unitDiff} $ \\
    \hline
    \hline
    \end{tabular}
    \label{tab:InitialParametersHC}
\end{table}

\paragraph{Parameter studies}
To enable computationally feasible simulation times, the evaporation rate is artificially scaled by a factor of $5.0\cdot 10^{4}$, following the approach proposed by Wolf~\cite{wolf2024computational} and adopted in our previous work~\cite{weichel2025modeling}.
To preserve the capillary number, the solvent viscosity is reduced by the same factor.
Through this approach, the computational cost can be significantly reduced.
Furthermore, the diffusion coefficient is scaled by the same factor to maintain the P\'eclet  number constant, thereby preventing an under representation of diffusion. 
The binder distribution is evaluated using the previously defined binder concentration field $\fieldresult$ (see Eq.~\ref{eq:fieldresult}).
% \begin{align} 
% \fieldresult = \field I + (1-I)\fielda. 
% \end{align} 
To quantify the binder distribution after electrode drying, the concentration $\fieldresult$ is evaluated using the following procedure. 
First, horizontal averaging over the $x$-direction is applied
\begin{align}
    \xAverage{\fieldresult} = \frac{1}{l_{\text{x}}} \intX{0}{l_{\text{x}}}{\fieldresult},
\end{align}
yielding a  one dimensional vertical concentration profile along the $y$-direction at the reference time $t_0$ and the final time $t_{\text{end}}$. 
The resulting profiles are truncated at the last location where the concentration exceeds a negligible threshold.
Subsequently, the relative changes in the binder distribution 
\begin{align}
    R(y) = \frac{\xAverage{\fieldresult}(y,t_{\text{end}})}{\xAverage{\fieldresult}(y,t_0)}
\end{align}
are computed. 
A linear regression of $ R(y)$ yields the slope $\steigungGradient$, which defines the normalized vertical binder gradient.
This scalar quantity is used as the primary metric for comparing binder migration across different simulations.
In a first study, the impact of the parameters contributing to the capillary number 
\begin{align} 
\Ca \, {=} \,\frac{\mu_{\text{solvent}} \charVelocity}{\sigma}, 
\end{align} 
on the final binder distribution is examined by varying the solvent viscosity $\mu_{\text{solvent}}$, the evaporation rate $\Dot{r}$, and the surface tension $\sigma$.
An increased evaporation rate leads to a steeper binder gradient, whereas a higher viscosity counteract this effect. 
Additionally, the effect of changing the contact angle is analyzed while changing the viscosity. 
Finally, the breakthrough time $t^{\text{breakthrough}}$, defined as the moment when the air phase first reaches the substrate, is analyzed as an additional characteristic parameter.
\subsection{Temporal evolution of drying}
\begin{figure}
\centering
\includegraphics[]{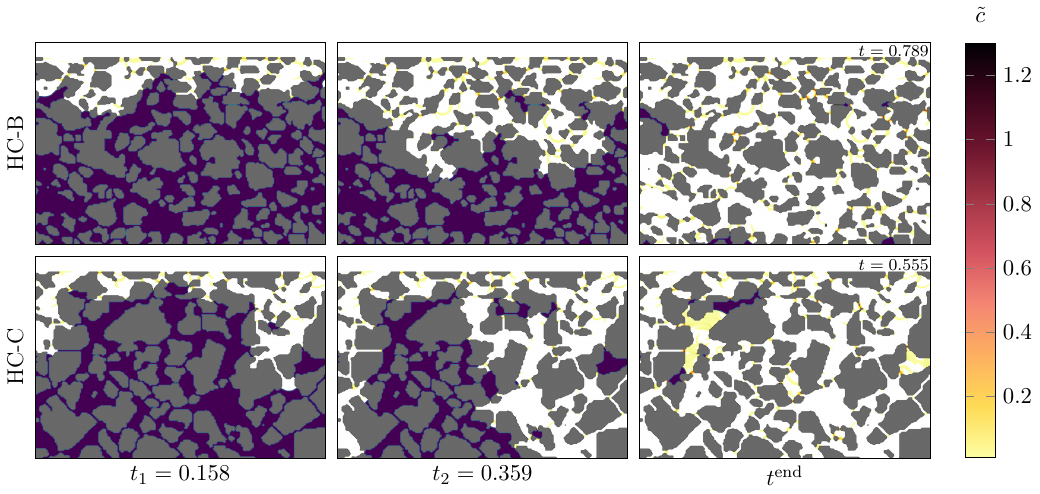}
\caption{Temporal evolution of the solvent and binder concentrations for HC-B (top row) and HC-C (bottom row). The first two temporal sequences correspond to the identical  nondimensionalized  times $t_1= 0.158$ and $t_2=0.287 $, whereas the last pictures correspond to the end of drying for the  nondimensionalized  times $t_{\text{HC-B}}^{\text{end}}=0.789$ and  $t_{\text{HC-C}}^{\text{end}}=0.555$.  The normalized vertical binder gradients are $\steigungGradient = 0.61$ and $\steigungGradient = 1.45$ for HC-B and HC-C, respectively. The solvent is depicted in blue and the particles in grey. For clarity, the binder distribution is displayed only for concentrations larger than $c > 0.01$.}
\label{fig:HC-Time-evolution}
\end{figure}

Figure~\ref{fig:HC-Time-evolution} depicts the distribution of the solvent and the binder~$\fieldresult$ for the two microstructures  at selected nondimensionalized times for a capillary number of $\Ca = 6.23 \cdot 10^{-3}$. 
HC-B is illustrated in the upper row and HC-C in the lower row. 
The first two columns correspond to the nondimensionalized times $t_1= 0.158$ and $t_2=0.359 $, while the last column shows the state at the end of drying ($t_{\text{HC-B}}^{\text{end}}=0.789$ and  $t_{\text{HC-C}}^{\text{end}}=0.555$).
Only regions exceeding the initial value of $c=0.01$ are displayed for the binder distribution.
The simulations for both microstructures demonstrate that binder accumulates in areas where the interparticle distance is relatively small and capillary channels are narrow.
% Such narrow capillary channels are preferred for the solvent flow, since larger pores drain prior smaller ones according to the Laplace pressure, and so promote a stronger binder accumulation. 
Such narrow capillary channels are preferential pathways for solvent flow, as larger pores drain prior to smaller ones due to differences in Laplace pressure, thereby promoting enhanced binder accumulation.
% This behavior results from the pore emptying sequence governed by Laplace pressure, where larger pores drain prior to smaller ones.
% Consequently, solvent flow is funneled through narrow capillary channels, promoting local binder accumulation.
A comparison of  the two microstructures reveals that the microstructure HC-B exhibits a more homogeneous binder distribution  over the electrode height, whereas the binder distribution in the  HC-C microstructure is predominantly concentrated in the upper region. 
This trend is also reflected in the evaluation of the normalized vertical binder gradient $\steigungGradient$. 
For comparison, HC-B exhibits a normalized vertical binder gradient of $\steigungGradient = 0.61$, whereas HC-C shows a higher value of $\steigungGradient = 1.45$.
These findings are consistent with experimental observations~\cite{klemens2023challenges}, which report reduced binder migration in microstructures composed of smaller particles and narrower pore networks.
%%%%%%%%%%%%%%%%%%%%%%%%%%%%%%%%%%%%%%%%%%%%%%%
%%%%%%%%%%%%Zeitverlauf plotten%%%%%%%%%%%%%%%%
%%%%%%%%%%%%%%%%%%%%%%%%%%%%%%%%%%%%%%%%%%%%%%%
%%%%%%%%%%%%%%%%%%%%%%%%%%%%%%%%%%%%%%%%%%%%%%%
%%%%%%%%%%%%Zeitverlauf plotten%%%%%%%%%%%%%%%%
%%%%%%%%%%%%%%%%%%%%%%%%%%%%%%%%%%%%%%%%%%%%%%%
\subsection{Modification of the solvent viscosity}
\begin{figure}[htbp]
\centering
\includegraphics[]{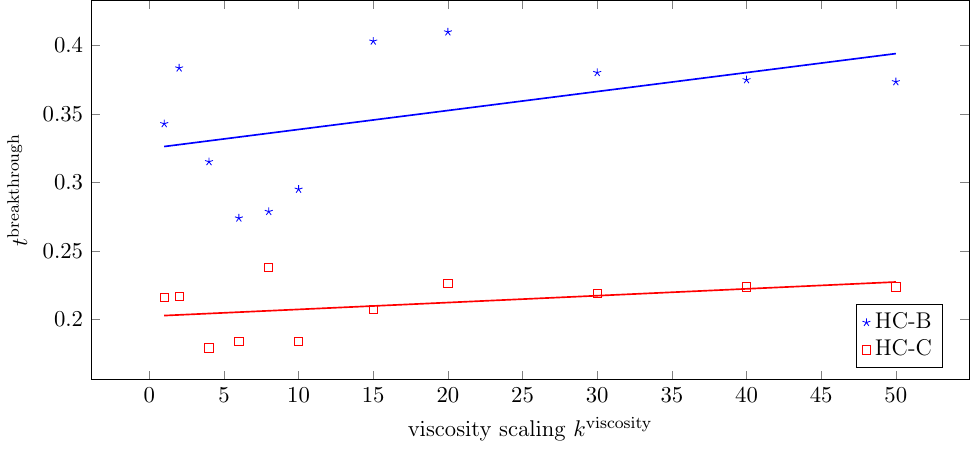}
\caption{Nondimensionalized breakthrough times as a function of the solvent viscosity $\mu_{\text{solvent}} \, {=} \facvis\mu_{\text{solvent},0}$. The results for microstructure HC-B and HC-C are shown in blue and red, respectively.}  
\label{fig:capillary-vs-time}
\end{figure}

\begin{figure}
\centering
\includegraphics[]{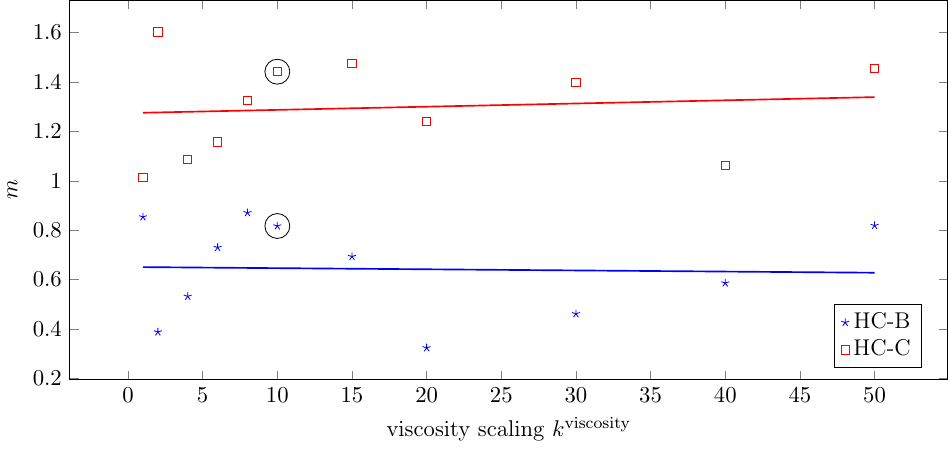}
\caption{Normalizied binder gradient $\steigungGradient$ as  a function of the solvent viscosity $\mu_{\text{solvent}} \, {=} \facvis\mu_{\text{solvent},0}$. Black circles indicate the reference case for the further simulation studies, displayed in Figure~\ref{fig:variation-evap} and~\ref{fig:variation-sigma} with varying evaporation rate and  surface tension, respectively.}
\label{fig:gradient-vs-mu}
\end{figure}

To further investigate the impact of the solvent viscosity  on the final binder distribution, this section examines the effect of increasing solvent viscosity under  constant evaporation rate and surface tension. 
Thereby, the reference viscosity $\mu_{\text{solvent},0}$ is varied by the factor $\facvis$, with $\facvis \in \{1, 2, 4, 6, 8, 10, 15, 20, 30, 40, 50\}$.
Figure~\ref{fig:capillary-vs-time} shows the nondimensionalized breakthrough time $t^{\text{breakthrough}}$ as a function of the solvent viscosity, while figure~\ref{fig:gradient-vs-mu} presents the vertical binder gradient $\steigungGradient$ as a function of solvent viscosity.
The results are shown in blue and red for the microstructures HC-B and HC-C, respectively.
For both microstructures, the breakthrough time increases with increasing viscosity, exhibiting a behavior consistent with previous studies~\cite{metzger2008viscous,weichel2025modeling}.
% Notably, the breakthrough times obtained for the microstructure HC-B are higher than those for HC-C, which contrasts with our previous findings~\cite{weichel2025modeling}.
% This discrepancy may be attributed to differences in the initial microstructures used in the simulations, in combination with the different values chosen for $\kappa$.
With respect to binder migration, microstructure HC-C exhibits a larger vertical gradient than microstructure HC-B  for all considered viscosities, which can be explained by the smaller particle size of HC-B~\cite{klemens2023process}.
Furthermore, neither microstructure shows a systematic increase in the binder gradient with increasing viscosity, indicating that viscosity alone does not constitute a dominant control parameter for binder migration under the investigated drying conditions.
It should be noted that real slurries exhibit shear-thinning behavior~\cite{klemens2023process} and sometimes yield point behavior~\cite{burger2024additive}, which is not captured by the present model~\cite{weichel2025modeling}, but can have an influence.

\subsection{Modification of the evaporation rate}
\begin{figure}[htbp]
\centering
\includegraphics[]{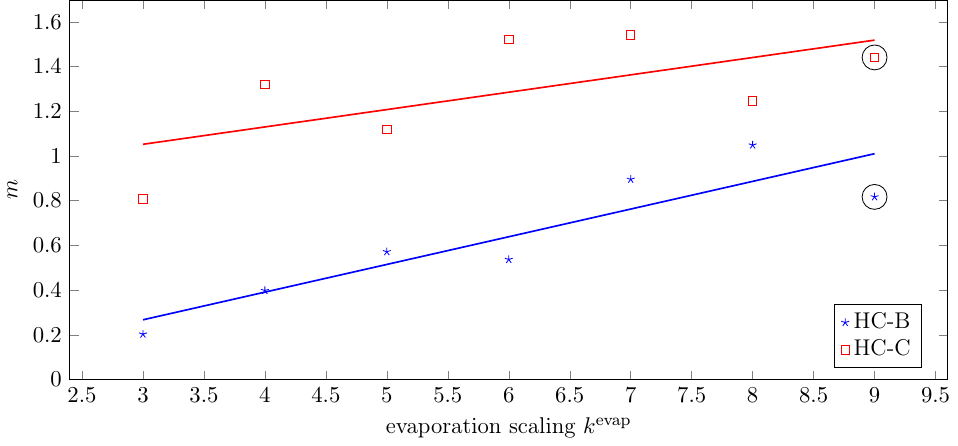}
\caption{Normalized slope of the binder gradient $\steigungGradient$ as a function of the evaporation rate.
The evaporation rate is given by $\Dot{r} = \facevap \Dot{r}_0$, and  $\Dot{r}_0 = \SI{1}{\gram\per\square\meter\per\second}$.
The results for microstructures HC-B and HC-C are shown in blue and red, respectively.
A viscosity of $\mu = 10,\mu_0$ is used in the simulation study.
The black circle indicate the marked cases from the previous section.}
\label{fig:variation-evap}
\end{figure}

To increase the throughput of dried electrodes, the evaporation rate can be enhanced.
However, higher drying rates lead to stronger binder gradients across the electrode thickness, which can deteriorate the chemical and mechanical properties of the electrode~\cite{klemens2023process,burger2024additive}.
This section investigates the influence of the evaporation rate and the microstructure on the gradients in the final binder distribution. 
For this purpose, the evaporation rate is varied according to $\Dot{r} = \facevap\Dot{r}_0$ with $\facevap \in \{3, 4, 5, 6, 7, 8, 9\}$ and $\Dot{r}_0 = \SI{1}{\gram\per\square\meter\per\second}$. 
The viscosity is fixed at $\mu = 10\mu_0$ for all simulations.
Figure~\ref{fig:variation-evap} displays the results for the microstructures HC-B and HC-C in blue and red, respectively. 
The black circles indicate the cases marked in black from the previous section corresponding to a factor of $\facevap = 9$.
For all simulations, the normalized gradient $\steigungGradient$ is higher for the microstructure HC-C. 
This can be attributed to the larger particle diameter of HC-C, which is consistent with findings reported in the literature~\cite{klemens2023process} and with the discussion in the previous section.
Furthermore, it is  observed that the binder gradient increases with increasing evaporation rate for all investigated drying conditions. 
For the microstructure HC-C this trend is in good agreement with previous experimental studies~\cite{jaiser2017development, klemens2023process, burger2024additive}.
A similar qualitative trend is observed for the HC-B microstructure in the present work.
However, this finding is not fully consistent with the results reported by Klemens et al.~\cite{klemens2023process}. 
In their study, hard carbon electrodes with smaller particle sizes exhibit a constant or even increased  adhesion force with increasing evaporation rate, indicating a less pronounced binder migration. 
This behavior is attributed to a reduced amount of binder that can be transported by capillary driven flow in systems with smaller particles. 
In such microstructures, the slurry rheology exhibits a gel-like CMC/SBR network, which immobilizes a significant fraction of the binder and counteracts the capillary forces during the pore emptying process.
In contrast, the present model predicts  an increasing binder gradient for the HC-B microstructure with increasing drying rate. 
This result is consistent with the underlying physics of the model, in which higher evaporation rates lead to stronger capillary forces, while rheological effects associated with shear-thinning are neglected.
Consequently, although  the model correctly captures the fundamental capillary driven mechanism of binder migration, it overestimates binder mobility for microstructures characterized by small particle sizes.
This discrepancy highlights the importance of incorporating shear-dependent viscosity  into future model extensions in order to accurately represent binder transport in battery electrodes across different particle size regimes.

\subsection{Variation of the surface tension}
\begin{figure}[htbp]
\centering
\includegraphics[]{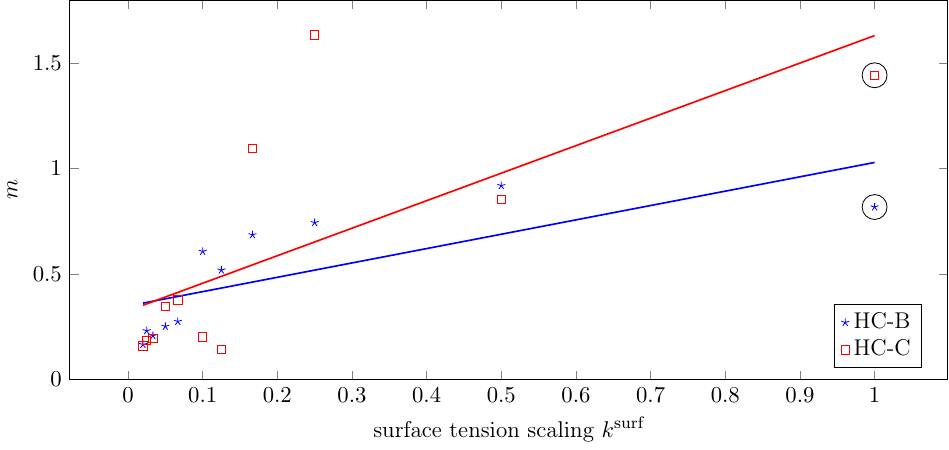}
\caption{Resulting normalized binder gradient $\steigungGradient$ for both microstructures.
The HC-B microstructure is shown in blue, while HC-C is shown in red.
The surface tension $\sigma_0$ is varied by a factor of $\facsigma$ according to $\sigma = \facsigma\sigma_0$. 
The viscosity is set to $\mu = 10\mu_0$ for all simulations with the reference case marked with the black circles.
}
\label{fig:variation-sigma}
\end{figure}

\begin{figure}[htbp]
\centering
\includegraphics[]{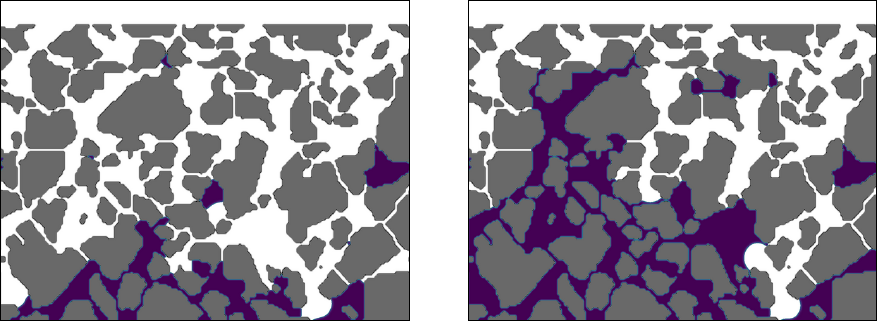}
\caption{Solvent distribution at the time of breakthrough for $\facsigma = 1/8$  (left) and $\facsigma = 1/4$ (right) for microstructure HC-C. The resulting slopes of the binder gradient $\steigungGradient$ are $\steigungGradient = 1.63$ and $\steigungGradient = 0.14$ for $\facsigma = 1/4$ and $\facsigma = 1/8$, respectively.}
\label{fig:sigma-solventfront}
\end{figure}

In this section the capillary number $\Ca$ is varied by modifying the surface tension $\sigma$ according to $\sigma = \facsigma\sigma_0$. 
The viscosity is set to $\mu = 10\mu_0$ for all simulations and an evaporation rate of $\Dot{r} = \SI{9}{\gram\per\square\meter\per\second}$ is chosen.
% Figure~\ref{fig:variation-sigma} presents the results of the simulation study for $\facsigma \in \{1, 2, 4, 6, 8, 10, 15, 20, 30, 40, 50\}$. 
Figure~\ref{fig:variation-sigma} presents the results of the simulation study for $\facsigma \in \{1/50, 1/40, 1/30, 1/20, 1/15, 1/10, 1/8, 1/6, 1/4, 1/2, 1\}$. 
The results for microstructure HC-B are shown in blue, while those for microstructure HC-C are shown in red.
For both microstructures, the normalized binder gradient $\steigungGradient$ increases with increasing factor $\facsigma$, which indicates that a reduced surface tension leads to a weaker binder accumulation at the surface of the electrode. 
The impact of the surface tension variation is more pronounced for microstructure HC-C, whereas for smaller values of  $\facsigma$ the gradients of both microstructures become  nearly identical, suggesting a reduced influence of the chosen microstructure in this regime.
Furthermore, for microstructure HC-C a pronounced drop in the normalized binder gradient $\steigungGradient$ is observed  between $\facsigma = 1/4$ and $\facsigma= 1/8$. 
The overall decrease in the normalized binder gradient with a lower surface tension reduction factor $\facsigma$ can be attributed to weakened capillary forces and thus more homogeneous drying front.
As a result, upward solvent flow by capillary driven convection and thus binder transport towards the electrode surface are suppressed.
The sharp increase of the normalized binder gradient observed for HC-C between  $\facsigma = 1/8$ and $\facsigma= 1/4$ can be explained by differences in the solvent front at the time of breakthrough to the substrate. 
These patterns are displayed in figure~\ref{fig:sigma-solventfront}. 
On the left, the solvent front for a factor of $\facsigma=1/8$ is illustrated, while the result for a factor of $\facsigma = 1/4$ is shown on the right. 
For the lower surface tension ($\facsigma=1/8$) the drying front appears homogeneous compared with the case of higher surface tension ($\facsigma=1/4$), where pronounced pinning at elevated positions within the microstructure is observed.
Due to this pinning, the binder is transported toward the electrode surface over a longer period, driven by sustained solvent flow.
These findings highlight the importance of physically consistent modeling approaches for binder migration.
In particular, a simple diffusion–convection equation that neglects capillary effects is insufficient to accurately describe drying induced binder migration.
In addition, the study shows that binder migration can be influenced by a change in surface tension.

\subsection{Variation of the contact angle}
\begin{figure}[htbp]
\centering
\includegraphics[]{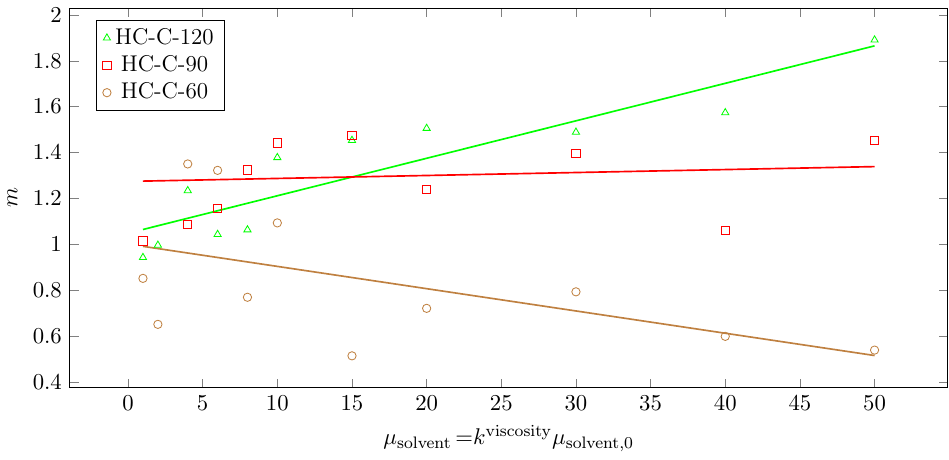}
\caption{
Normalized gradient $\steigungGradient$ as functions of the viscosity for HC-C with different angles (\ang{120}, \ang{90}, \ang{60}).
}
\label{fig:HC-C-combined-angles}
\end{figure}

In this section the impact of the contact angles, \ang{90}, \ang{60} and \ang{120}, on the microstructure HC-C is investigated.
The simulations are performed using the reference parameter set listed in table~\ref{tab:InitialParametersHC}.
In this study, a contact angle of~\ang{60} represents hydrophobic behavior, whereas a contact angle of~\ang{120} corresponds to hydrophilic behavior~\cite{weichel2025modeling}.
Figure~\ref{fig:HC-C-combined-angles} illustrates the influence of solvent viscosity on the binder gradient $\steigungGradient$ for the different contact angles.
The results show that, for both hydrophilic and hydrophobic wetting conditions, the sensitivity of the binder gradient to changes in viscosity is more pronounced compared with the reference case of~\ang{90}.
In the hydrophilic case, the binder gradient increases more strongly with increasing viscosity, although it does not necessarily reach the highest absolute value for all viscosities considered.
In contrast, hydrophobic wetting behavior exhibits a different trend, where  the binder gradient remains less pronounced with an increasing viscosity.
In summary, deviations of the contact angle from \ang{90} significantly amplify the sensitivity of binder migration to solvent viscosity.
This demonstrates that wettability of the active material constitutes a control parameter for tailoring binder migration in battery electrodes.
\section{Summary and outlook}
\label{sec:summary}
\paragraph{Summary} Most studies investigating binder migration during the drying of battery electrodes rely on simplified one-dimensional (1D) models. 
In these approaches, binder migration is commonly attributed to film shrinkage. 
However, according to the current state of knowledge, this assumption is incomplete, as binder migration predominantly occurs due to capillary transport during pore emptying.
In this work, a diffusion–convection equation is coupled with a fully resolved pore emptying model to describe binder migration driven by capillary transport. 
The newly developed model is validated using two validation cases.
First, the concentration distribution in a bubble-rise benchmark is analyzed, demonstrating that, in the absence of deposition, the concentration is transported by convection within the correct phase. 
Second, drying without deposition is investigated using a simplified setup by evaluating two Péclet numbers. 
The results are in good agreement with previously reported findings on binder migration.
Subsequently, the model is applied to the drying process of real hard carbon electrodes. 
The influence of microstructure, viscosity, evaporation rate, surface tension, and contact angle is investigated. 
The considered microstructures differ in their pore size distributions and mean particle diameters.
The simulations reveal that microstructures with smaller particle diameters generally exhibit a more homogeneous binder distribution. 
This behavior is consistent with observations reported in the literature \cite{klemens2023process}. 
With respect to variations in viscosity, the simulations indicate only a comparatively minor influence on binder distribution.
A more pronounced impact regarding the viscosity can be seen by altering the wetting behavior. 
In contrast, increasing evaporation rates lead to a more pronounced binder inhomogeneity.
A similar trend is observed with increasing surface tension.
Overall, these results indicate that binder homogeneity can be effectively optimized by adjusting the contact angle and surface tension, thereby enabling an improved control of the drying process as a whole.
In addition, this study demonstrates that models aiming to predict binder migration during battery electrode drying should explicitly incorporate capillary forces and microstructural information. \\
\paragraph{Outlook} In future works, the model is planned to be refined further to depict transport processes during pore emptying more accurately. In the present work, two-dimensional (2D) simulations were considered, which inherently simplify the pore size distribution and its impact on drying behavior and consequently on binder migration. Future work will focus on fully resolved three-dimensional (3D) microstructures to assess the influence of this simplification. Achieving this will require detailed structural data, which are currently limited.
In addition, the influence of mass transport resistance in the gas phase will be integrated into the model, since saturation of the gas phase in the pores may amplify capillary driven transport towards the top by mitigating evaporation inside the pore structure. This is especially true for thicker electrodes, for which also the assumption of a linear drying rate becomes increasingly inaccurate \cite{kumberg2021influence}. 
Furthermore, the model will be extended to account for shear-thinning solvent behavior, enabling a more realistic representation of non-Newtonian rheology during the drying process.

\section*{CRediT authorship contribution statement}
\textbf{Marcel Weichel:} Methodology, Software, Validation, Investigation, Visualization, Writing – original draft, Writing – review \& editing. 
\textbf{Martin Reder:}  Investigation, Writing – review \& editing, Supervision.
\textbf{Gerit Mühlberg:}  Software, Writing – review \& editing.
\textbf{David Burger:}  Investigation, Writing – review \& editing.
\textbf{Philip Scharfer:}  Writing – review \& editing.
\textbf{Wilhelm Schabel:}  Writing – review \& editing.
\textbf{Britta Nestler:}  Writing – review \& editing, Supervision, Funding acquisition.
\textbf{Daniel Schneider:}  Writing – review \& editing, Supervision, Funding acquisition.
\section*{Declaration of Generative AI and AI-assisted technologies in the writing process}
The authors used DeepL (DeepL SE) and ChatGPT (OpenAI) to improve readability and clarity. All tool-assisted text was subsequently reviewed and edited by the authors, who take full responsibility for the final content of the publication.
\section*{Data availability}
The raw data underlying the plots, as well as additional data, are provided as supplementary material.
Further data supporting the findings of this study are available from the corresponding author upon reasonable request.
The related simulation software is commercially available; 
further information can be found at \url{https://www.iam.kit.edu/mms/pace3d.php}.
\section*{Acknowledgments}
This work contributes to the research performed at CELEST (Center for
Electrochemical Energy Storage Ulm-Karlsruhe) and was funded by the
German Research Foundation (DFG) under Project ID 390874152 (POLiS
Cluster of Excellence). Within the cooperation of the cluster we thank Dr. Marcus Müller (Institute for Applied Materials IAM-ESS) for support regarding the SEM measurements. Support regarding the model development is provided through funding by the Helmholtz association within the programme "MTET", no. 38.02.01, which is gratefully acknowledged.
Additionally, this research is funded by the Gottfried-Wilhelm Leibniz prize NE 822/31-1 of the German Research Foundation (DFG).

\appendix  
\section{Derivation of the binder equation}
\label{sec:binder-derivation}
Consider the boundary value problem
\begin{alignat}{3}
 \dot{\field} :=  \convGrad{\field} 
 &= \div (\diffusivity \grad \field) + \source,
 \qquad &&\pos \in \volume
 \label{EQ:pde_grad}\\
  (\diffusivity \grad \field) \cdot \normal &= \boundaryflux,
 &&\pos \in \partial\volume
 \label{EQ:bc}
\end{alignat}
Multiplying eq.~\eqref{EQ:pde_grad} with test function $\testFunc$ and integration over the volume~$\volume$ yields
\begin{align}
    0 &= \intVol{\volume}{
     -\testFunc \dot{\field} + \testFunc \div (\diffusivity \grad \field) + \testFunc\source 
    }  \\
   &= \intVol{\volume}{
     -\testFunc \dot{\field} +  \div (\testFunc \diffusivity \grad \field)
     - \diffusivity \grad \field \cdot \grad \testFunc + \testFunc\source 
    } \\
    &=  \intVol{\volume}{
     -\testFunc \dot{\field} 
     - \diffusivity \grad \field \cdot \grad \testFunc + \testFunc\source }
     + \intSurf{\partial \volume}{
     \testFunc \diffusivity \grad \field \cdot \normal
     }\\
     &= \intVol{\volume}{
     -\testFunc \dot{\field} 
     - \diffusivity \grad \field \cdot \grad \testFunc + \testFunc\source }
     + \intSurf{\partial \volume}{
     \testFunc \boundaryflux
     } \label{EQ:weak_form_grad}
\end{align}
This was done exploiting the Gauss divergence theorem and assuming the Neumann boundary conditions everywhere on~$\partial \volume$.
Consider a domain~$\domain$ with $\volume \subset \domain$ we exploit
\begin{align}
    \intVol{\volume}{(\cdot)} = \intVol{\domain}{\indicator[](\cdot)}
    \quad \text{and} \quad
    \intSurf{\partial\volume}{(\cdot)} = \intVol{\domain}{\SurfDirac(\cdot)}
\end{align}
with the indicator function $\indicator$ and the surface Dirac distribution~$\SurfDirac$, with $\grad \indicator = -\SurfDirac \, \normal$
The weak form~\eqref{EQ:weak_form_grad} becomes
\begin{align}
0 = \intVol{\domain}{
     -\testFunc \indicator \dot{\field} 
     - \indicator \diffusivity \grad \field \cdot \grad \testFunc + \testFunc \indicator \source
     + \testFunc \boundaryflux \, \SurfDirac
     }
\end{align}
Looking at the second summand, we obtain
\begin{align}
\intVol{\domain}{- \indicator \diffusivity \grad \field \cdot \grad \testFunc}
&=\intVol{\domain}{-  \div \left(\testFunc \indicator \diffusivity \grad \field \right) + \testFunc \div \left(\indicator \diffusivity\grad \field \right)}\\
&=\intVol{\domain}{\testFunc \div \left(\indicator \diffusivity\grad \field \right)}.
\end{align}
The last step uses the divergence theorem and exploiting $\forallHoldsText{\pos \in \partial \domain}\indicator=0$.
Therefore, we get
\begin{align}
0 &= \intVol{\domain}{
     -\testFunc \indicator \dot{\field} 
    + \testFunc  \div \left(\indicator \diffusivity\grad \field \right) + \testFunc \indicator \source
     + \testFunc \boundaryflux \, \SurfDirac
     }\\
&= \intVol{\domain}{ \testFunc \left[
     - \indicator \dot{\field} 
     + \div \left(\indicator \diffusivity\grad \field \right) + \indicator \source
     +  \boundaryflux \, \SurfDirac \right]
     }
\end{align}
Localising this gives
\begin{align}
    \indicator \dot{\field} = \div \left(\indicator \diffusivity\grad \field \right) + \indicator \source +  \boundaryflux \, \SurfDirac 
    \label{EQ:wholeDomain_grad}
\end{align}
using 
\begin{align}
\indicator \dot{\field} 
= \pdiffFrac{\field \indicator}{t} + \indicator \velocity \cdot \grad \field - \field \pdiffFrac{\indicator}{t}
\end{align}
we can re-write eq.~\eqref{EQ:wholeDomain_grad} as
\begin{align}
 \pdiffFrac{\field \indicator}{t} =- \indicator \velocity \cdot \grad \field + \div \left(\indicator \diffusivity\grad \field \right) + \indicator \source +  \boundaryflux \, \SurfDirac  + \field \pdiffFrac{\indicator}{t} 
 \label{eq:binder-equation}
\end{align}
% \include{00_Appendix_B}
% \include{00_Appendix_C}
% \include{00_Appendix_D}
% %\nocite{*}                    %Alle Referenzen aufnehmen
\bibliography{literatur_mw}      %Literaturverzeichnis

% \input{summary}
% \input{intro}
% \input{theory}
%\input{materialdata}
%\input{simulation-results-matlab}
% \input{simulation-results-pace}
%\appendix
%\input{appendix}

% \cite{nestler2005}

% \cite{Klemens2022}

% \printbibliography %Prints bibliography  
\end{document}